%% file: Paper_v2.tex
\documentclass{article}

\usepackage[preprint,nonatbib]{neurips_2025}
\usepackage[numbers,compress,sort]{natbib}

\usepackage[utf8]{inputenc}
\usepackage[T1]{fontenc}
\usepackage[colorlinks=true,linkcolor={blue!70!black},citecolor={blue!70!black},urlcolor={blue!70!black},hypertexnames=false]{hyperref}
\usepackage{url}
\usepackage{booktabs}
\usepackage{amsfonts}
\usepackage{nicefrac}
\usepackage{microtype}
\usepackage{xcolor}
\usepackage{graphicx}
\usepackage{amsmath,amssymb}
\usepackage{multirow}
\usepackage{longtable}
\usepackage[nomarkers,nofiglist,notablist]{endfloat}

\title{Rethinking the Choice Behavior of Sugar Metabolism in Bacteria\thanks{Repository: \url{https://github.com/varnerlab/Kompala-Model-LP-Paper.git}.}}

\author{%
  Jeffrey D. Varner \\
  R.F. Smith School of Chemical and Biomolecular Engineering \\
  Cornell University, Ithaca, NY 14853 \\
  \texttt{jdv27@cornell.edu} \\
}

\begin{document}

\maketitle

\begin{abstract}
\input{sections/abstract}
\end{abstract}

\section*{Introduction}
\input{sections/introduction}

\clearpage

\section*{Materials and Methods}\label{sec:methods}
\input{sections/methods}

\clearpage

\section*{Results}\label{sec:results}
\input{sections/results}

\clearpage

\section*{Discussion}\label{sec:discussion}
\input{sections/discussion}

\section*{Acknowledgments}
We gratefully acknowledge the suggestions from the anonymous reviewers to improve this manuscript.
Further, we thank the CHEME 5760 Quantitative Decisions course students for their feedback on the lecture
example that inspired this study. Finally, we thank the members of the Varnerlab for their input throughout this work.

\input{sections/backmatter}

\clearpage
\bibliographystyle{unsrtnat}
\bibliography{References_v1}

\clearpage
\appendix
\input{sections/supplement}

\end{document}

%% file: sections/abstract.tex
Ramkrishna, Kompala, and Tsao proposed the cybernetic model of microbial growth, in which cells allocate enzyme synthesis resources according to a matching rule that mimics rational decision-making. The matching rule was later shown to be optimal under general assumptions about the underlying return-on-investment structure, yet the specific objective the cell maximizes and the constraints bounding that choice were never written down as an explicit economic decision. Here we supply that missing decision, recasting cybernetic enzyme-synthesis control as a consumer choice problem from microeconomic theory: the cell allocates a limited proteome budget among competing catabolic enzymes as a linear program (LP), maximizing a linear growth utility subject to a linear proteome budget constraint. Because the utility is linear, the LP's solution is geometric: whenever the iso-utility line's slope differs from the budget constraint's, the optimum is a corner, and the entire proteome budget is allocated to the enzyme for the single most profitable substrate. Corner solutions correspond to diauxic growth, and sequential substrate consumption follows from the choice of corner rather than a distinct regulatory mechanism. Only when the two slopes coincide does the optimum spread across the entire budget line instead of concentrating at a single corner; this degenerate case is where simultaneous substrate use becomes admissible. Using kinetic parameters from single-substrate experiments together with a small set of model-level parameters set in this study, the LP-derived cybernetic variables reproduced the diauxic and triauxic batch growth of \emph{Klebsiella oxytoca} on glucose--xylose and glucose--xylose--lactose mixtures, achieving a fit comparable to the classical matching law. Thus, sequential substrate use is the generic outcome of growth-maximizing specialization under perfect substitutability, and co-utilization is the degenerate case of equal profitability.

%% file: sections/introduction.tex
When Jacques Monod grew \emph{Escherichia coli} on a mixture of glucose and lactose in the 1940s, the cells consumed glucose to exhaustion before switching to lactose, with a characteristic growth pause separating the two phases~\cite{Monod:1942}. He named this two-phase growth \emph{diauxie}, and its mechanism was later traced to catabolite repression, in which glucose suppresses transcription of the genes needed to metabolize alternative sugars~\cite{Deutscher:2008}. Catabolite repression is nonetheless only a proximate explanation, describing the molecular machinery that implements sequential substrate consumption without explaining why sequential consumption is the strategy a cell should adopt in the first place. Ramkrishna and colleagues addressed this deeper question through the cybernetic modeling framework, in which microorganisms behave as purposeful agents that allocate finite resources to maximize growth~\cite{Ramkrishna:1982,Dhurjati:1985aa,Kompala:1984aa,Kompala:1986aa}. The cell is governed by two control variables: $u_i$ regulating synthesis of the enzyme for substrate $i$, and $v_i$ regulating its activity. The synthesis allocation $u_i$ follows a matching rule that distributes resources in proportion to the instantaneous growth rate each substrate supports~\cite{Kompala:1986aa}. Built entirely from single-substrate kinetics, the resulting model reproduced the diauxic growth of \emph{Klebsiella oxytoca} on multiple sugar mixtures~\cite{Kompala:1986aa}. In a 1987 perspective, Ramkrishna, Kompala, and Tsao asked: \emph{are microbes optimal strategists?}~\cite{Ramkrishna:1987aa}

Over the following decades the cybernetic framework grew into a general, optimization-based description of metabolic dynamics, surveyed in a review and a recent monograph~\cite{Young:2015aa,RamkrishnaSong:2018aa}. Straight and Ramkrishna gave the framework a modular structure, decomposing a metabolic pathway into a small set of elementary structural motifs, each under its own cybernetic control, and used this construction to describe growth on complementary nutrients~\cite{Straight:1994aa}. Hybrid cybernetic models later placed the same control variables over a flux-mode decomposition of the metabolic network~\cite{Young:2008aa,Kim:2008aa}, extending the approach from lumped single-substrate uptake to genome-scale stoichiometry and reaching networks too large for an explicit elementary-mode basis~\cite{Vilkhovoy:2016aa}. Beyond batch diauxie, the models have described steady-state multiplicity, mutant phenotypes, and continuous-culture dynamics across a range of organisms~\cite{RamkrishnaSong:2018aa}. However, one theoretical question remains open: \emph{what do the cybernetic control laws optimize?} Young and Ramkrishna addressed this question, examining the optimality of the matching and proportional laws, deriving generalized forms of both, and showing that the matching law does not maximize average growth rate but instead reflects a frugal fitness-to-cost tradeoff~\cite{Young:2007aa}. Their optimal-control analysis established that the matching rule is optimal under specific assumptions about the return-on-investment structure of the problem, yet it left unidentified what the cell actually maximizes. Neither the utility being optimized nor the constraint bounding it was written as an explicit program with a geometric solution. Thus, the matching rule has remained a control heuristic with proven optimality properties rather than the solution of a stated economic decision. That decision itself has not been written down.

In this study, we recast cybernetic enzyme allocation as the consumer choice problem of microeconomic theory: a linear program (LP) that maximizes a linear growth utility subject to a linear proteome budget~\cite{Varian:1992}. Because a rational agent maximizing a linear utility under a budget spends it entirely on the good with the highest utility per unit cost, the optimum is a corner of the budget simplex. For a cell, this means exclusive synthesis of the enzyme responsible for consuming the single most profitable substrate. Diauxie is therefore the outcome of rational specialization under a finite budget with linear preferences. The diauxic lag is the time required to redirect synthesis after the LP switches corners at substrate exhaustion. Solving this LP at each integration step to obtain the cybernetic $u$-variables, we integrated the governing ODEs for \emph{K.~oxytoca} on glucose--xylose (diauxic) and glucose--xylose--lactose (triauxic) mixtures, with all parameters fixed from single-substrate experiments~\cite{Kompala:1986aa}. The model quantitatively reproduced the measured cellmass trajectories, the sequential substrate-depletion profiles, and the order of substrate preference. The $u$-variable time series showed step-function switching between corners, with no intermediate allocation at any step. Co-utilization appeared in the same framework as the degenerate case, arising when two substrates yield equal growth per unit proteome expenditure (the budget line is parallel to the indifference curve). Thus, diauxie is growth-maximizing specialization under perfect substitutability, and the cell's substrate choice is the corner solution of a budget-constrained utility maximization.

%% file: sections/methods.tex
\subsection*{Formulation of model equations.}
The model consists of $2n+1$ ordinary differential equations (ODEs) that describe the dynamics of the $n$ sugar concentrations,
the abundances of the $n$ pseudo enzymes that consume them, and the biomass concentration.
We discretize the continuous material balance equations describing carbon-limited batch growth on sugar set $\mathcal{S}$ using a forward difference
approximation with fixed step-size $h$ to obtain the discrete system (where $i$ denotes the sugar index, and $j$ denotes the time index):
\begin{eqnarray}
s^{(i)}_{j} & = & s^{(i)}_{j-1} - h\cdot{\left(\frac{r_{i,j-1}}{Y_{i}}\right)}\cdot{v}_{i,j-1}\cdot{c_{j-1}} \quad{\forall{i}\in\mathcal{S}} \label{eqn:substrate-mat-bal} \\
e_{j}^{(i)} & = & e_{j-1}^{(i)} + h\cdot{\left(\frac{u_{i,j-1}}{\tau_{i}} - \left(\mu_{j-1}+\beta_{i}\right)\cdot{e^{(i)}_{j-1}}+\lambda_{i}\right)}\quad{\forall{i}\in\mathcal{S}} \label{eqn:enzyme-mat-bal} \\
c_{j} & = & c_{j-1} + h\cdot{\left(\sum_{i\in\mathcal{S}}r_{i,j-1}\cdot{v_{i,j-1}} - k_{d}^{c}\right)}\cdot{c_{j-1}} \label{eqn:cell-mat-bal}
\end{eqnarray}
The quantity $s^{(i)}_{j}$ denotes the concentration of sugar $i$ at time $t_{j}$, $e_{j}^{(i)}$ denotes the concentration of pseudo enzyme $i$ at time $t_{j}$,
and $c_{j}$ denotes the concentration of cells at time $t_{j}$.
The parameter $\tau_{i}$ denotes the characteristic time constant for the expression of pseudo enzyme $i$.
A common value $\tau = 0.623$ h was used for all enzymes in this study.
The $\lambda_{i}$ parameter denotes the rate of unregulated (constitutive) pseudo enzyme $i$ synthesis.
The quantity $\mu_{j-1}$ denotes the specific growth rate of the culture at time $t_{j-1}$, a single rate shared by every pseudo enzyme balance and equal to the biomass growth term in the cell balance.
The parameter $\beta_{i}$ denotes the degradation rate constant for pseudo enzyme $i$.
The parameter $k_{d}^{c}$ denotes the cell death rate constant (units: $\mathrm{h}^{-1}$), which is distinct from the enzyme degradation rate $\beta_{i}$. Here $k_{d}^{c} = 0.022~\mathrm{h}^{-1}$.
The yield coefficient $Y_{i}$ denotes the biomass produced per unit of sugar $i$ consumed (units: gDW/g), consistent with the factor $1/Y_{i}$ in the substrate balance, Eqn.~\eqref{eqn:substrate-mat-bal}.

The $r_{i,j-1}$ terms denote the rate of consumption of sugar $i$ at time $t_{j-1}$ for growth,
where $r_{i,j-1}$ follows the Monod-like expression:
\begin{equation}\label{eqn:monod-like-rate}
r_{i,j-1} = \mu_{i}^{\text{max}}\cdot\left(\frac{e^{(i)}_{j-1}}{e_{i}^{\text{max}}}\right)\cdot\left(\frac{s^{(i)}_{j-1}}{K_{i}+s^{(i)}_{j-1}}\right)\quad{\forall{i}\in\mathcal{S}}
\end{equation}
The $\mu_{i}^{\text{max}}$ parameter denotes the maximum specific growth rate of cells on sugar $i$,
$s^{(i)}_{j-1}$ is the concentration of sugar $i$ one time step earlier,
and $K_{i}$ denotes the growth saturation constant for sugar $i$.
The $e^{(i)}_{j-1}/e_{i}^{\text{max}}$ term denotes the scaled abundance of pseudo enzyme $i$ at time $t_{j-1}$,
where $e_{i}^{\text{max}}$ is a fixed reference abundance that normalizes pseudo enzyme $i$ (a scaling constant rather than an enforced upper bound; the dynamics do not constrain the abundance to lie below it, and it can exceed $e_{i}^{\text{max}}$ during rapid induction). Because enzyme abundances are scaled by this reference, $e_{i}^{\text{max}}$ is dimensionless.
The expression of the pseudo enzyme $i$ depends upon some hypothetical scarce resource $R$,
a cybernetic control variable $0\leq{u_{i,j-1}}\leq{1}$ and other parameters.
Finally, the terms $v_{i,j-1}$ denote the cybernetic variable which controls
the activity of the pseudo enzyme $i$ at time $t_{j-1}$, i.e., $0\leq{v_{i,j-1}}\leq{1}$.
The specific growth rate that dilutes the pseudo enzymes is the activity-weighted sum of the substrate growth rates, $\mu_{j-1} = \sum_{i\in\mathcal{S}} r_{i,j-1}\,v_{i,j-1}$.
We list the parameter values used in this study (Table~\ref{tbl:parameters}).

\subsection*{Formulation of the cybernetic control variables.}
The cybernetic v-variable $v_{i,j}$, which controls the activity of the pseudo enzyme $e_{i}(R)$ at time $t_{j}$,
is defined using the cybernetic proportional law:
\begin{equation}
v_{i,j} = \frac{r_{i,j}}{\max_{k}r_{k,j}}\quad{\forall{i}\in\mathcal{S}}
\end{equation}
where the denominator $\max_{k}r_{k,j}$ computes the maximum rate of sugar consumption at time $t_{j}$.
Thus, the cybernetic v-variable $v_{i,j}$ is a normalized measure of the rate of sugar consumption by the pseudo enzyme $e_{i}(R)$ at time $t_{j}$, and
it is bounded by $0\leq{v_{i,j}}\leq{1}$.
See Kompala et al. and Young and Ramkrishna for a discussion of the proportional law, and the derivation of $v_{i,j}$ \cite{Kompala:1986aa, Young:2007aa}.

The cybernetic u-variable $u_{i,j}$, which controls the expression of pseudo enzyme $i$ at time $t_{j}$,
represents the fraction of total hypothetical resource $R$ allocated to the expression of enzyme $i$ at time $t_{j}$.
Traditionally, the value of the u-variables relied on the matching law formulation of the cybernetic control problem (see Young and Ramkrishna \cite{Young:2007aa}).
However, in this study, we explore a different approach: at each time step the cell chooses a \emph{target} allocation of pseudo enzyme abundances that maximizes a growth utility subject to a resource constraint on their expression.
We leave the exact nature of the resource unspecified, e.g., ATP or ribosomal capacity, but assume it is scarce, so the cell has limited resources to allocate.
Thus, the cell maximizes its utility by prudently allocating scarce resources among competing alternative pseudo enzymes.

We distinguish two abundances for each pseudo enzyme. The dynamic abundance $e^{(i)}_{j}$ is the state variable that appears in the growth rate $r_{i,j}$, Eqn.~\eqref{eqn:monod-like-rate}, and evolves by the enzyme balance, Eqn.~\eqref{eqn:enzyme-mat-bal}. The target abundance $\hat{e}_{i}$ is the decision variable of the allocation problem: the abundance the cell would hold under the current substrate concentrations if it could re-tool synthesis instantaneously. The allocation problem is solved for $\hat{e}_{i}$; its solution sets the synthesis control $u_{i,j}$, which then drives the dynamic abundance $e^{(i)}_{j}$ toward the target through Eqn.~\eqref{eqn:enzyme-mat-bal}. We label glucose as index $1$ and the other sugars as indices $2,\dotsc,n$.
Let $\mathcal{E} = \left\{e_{1}(R),\dotsc,e_{n}(R)\right\}$ denote the collection of pseudo enzymes that consume the sugars $\mathcal{S}$, with target abundances $\hat{e} = (\hat{e}_{1},\dotsc,\hat{e}_{n})$.
We assume a growth-utility function $U:\hat{e}\rightarrow\mathbb{R}$ of the form:
\begin{equation}\label{eqn:growth-utility}
U(\hat{e}_{1},\dotsc,\hat{e}_{n}) = \sum_{i\in\mathcal{S}}\gamma_{i}\,\hat{e}_{i}
\end{equation}
where $\gamma_{i}$ is the marginal growth return per unit abundance of pseudo enzyme $i$ at the current substrate concentration,
\begin{equation}\label{eqn:gamma-def}
\gamma_{i} \equiv \left(\frac{\mu_{i}^{\text{max}}}{e_{i}^{\text{max}}}\right)\cdot\left(\frac{s_{i}}{K_{i}+s_{i}}\right).
\end{equation}
The coefficient $\gamma_{i}$ collects the state-dependent factors of the Monod-like rate, so evaluating it at the dynamic abundance returns the realized growth rate, $\gamma_{i}e^{(i)}_{j} = r_{i,j}$, while $U$ in Eqn.~\eqref{eqn:growth-utility} is the \emph{potential} growth return the cell would realize if abundances were at the target $\hat{e}$. The two coincide only when the dynamic abundance equals the target; in general the allocation problem maximizes the potential return per unit budget and thereby sets the direction of synthesis, and realized growth follows as the dynamic abundance accumulates.
The expression of pseudo enzyme $e_{i}(R)$ is limited by the availability of a hypothetical resource $R$. Thus, we write a resource constraint of the form:
\begin{equation}\label{eqn:resource-constraint}
\sum_{i\in\mathcal{S}} b_{i}\cdot\hat{e}_{i} = B
\end{equation}
where $b_{i}$ denotes the resource $R$ consumed by the unit expression of pseudo enzyme $e_{i}(R)$,
and $B$ denotes the total amount of resource $R$, which we assume is constant.
The quantity $B$ represents the cell's finite capacity for enzyme synthesis, the proteome fraction available to the catabolic enzymes. It is normalized away in Eqn.~\eqref{eqn:resource-constraint-scaled}, so only the relative costs $\hat{b}_{i}$ enter.
While Eqn.~\eqref{eqn:resource-constraint} is correct in a mathematical sense, it is not a practical constraint
because it requires that we know the exact amount of resource $R$ consumed by the expression of each pseudo enzyme $e_{i}(R)$,
and the total amount of resource $R$ available to the cell. Instead, we divide both sides of Eqn.~\eqref{eqn:resource-constraint} by $B$ to obtain a fractional resource constraint:
\begin{equation}\label{eqn:resource-constraint-scaled}
\sum_{i\in\mathcal{S}} \hat{b}_{i}\cdot\hat{e}_{i} = 1
\end{equation}
where $\hat{b}_{i} = b_{i}/{B}$ denotes the fraction of resource $R$ consumed by the expression of pseudo enzyme $e_{i}(R)$.
Finally, we define the value of the control variable $u_{i,j}$ from the target abundance:
\begin{equation}
u_{i,j} \equiv \hat{b}_{i}\cdot\hat{e}_{i}
\end{equation}
which normalizes the allocation, $\sum_{i\in\mathcal{S}}u_{i,j} = 1$, placing it on the same allocation simplex on which the classical matching law is defined.
Putting these ideas together gives the problem the cell solves at each time step to determine which pseudo enzyme to express:
\begin{align*}
\text{maximize} &\quad \sum_{i\in\mathcal{S}}\gamma_{i}\,\hat{e}_{i} \\
\text{subject to} &\quad \sum_{i\in\mathcal{S}}\hat{b}_{i}\,\hat{e}_{i}\;\leq 1 \\
\text{and} & \quad 0 \leq \hat{e}_{i} \leq \hat{b}_{i}^{-1}\qquad{i=1,2,\dots,n}
\end{align*}
When at least one substrate has a strictly positive return ($\gamma_{i}>0$), the budget binds at the optimum, so the inequality and the balanced-growth equality $\sum_{i}\hat{b}_{i}\hat{e}_{i}=1$ coincide there. A substrate whose concentration falls below a depletion threshold has $\gamma_{i}=0$ (Eqn.~\eqref{eqn:gamma-def}) and drops out of the problem; if every return is zero the budget need not bind and the allocation is zero.
We define the profitability of substrate $i$, its growth return per unit of budget spent on the pseudo enzyme $e_{i}(R)$, as:
\begin{equation}\label{eqn:profitability-def}
\rho_{i} \equiv \frac{\gamma_{i}}{\hat{b}_{i}}.
\end{equation}

\subsection*{Geometry of the optimal resource allocation.}
The decision variables of this resource allocation problem are the target pseudo enzyme abundances $\hat{e}_{i}$, so the choice between expressing pseudo enzymes $i$ and $j$ for $i\neq{j}$ is the solution of a constrained linear utility maximization problem with a direct geometric reading (Fig.~\ref{fig:graphical-soln-ic-schematically}).
An indifference curve is a locus of constant utility, the combinations of the two goods that yield the same satisfaction. Its slope is the marginal rate of substitution, the rate at which the cell trades one good for another at fixed utility.
The resource constraint appears as the budget line, and the region beneath it is the set of allocations the cell can afford.
A higher-utility target such as point \texttt{a}, which would require expressing both pseudo enzymes, lies outside this region because the budget cannot afford it. The optimum is therefore forced onto the budget line itself.
Because the objective is linear, the outcome there is set by the slope of the indifference curves relative to the slope of the budget line.
An indifference curve is a level set of the linear utility $\gamma_{1}\hat{e}_{1}+\gamma_{2}\hat{e}_{2}$, and the budget line is the level set $\hat{b}_{1}\hat{e}_{1}+\hat{b}_{2}\hat{e}_{2}=1$; both slope downward, with magnitudes $m_{I}=\gamma_{1}/\gamma_{2}$ and $m_{B}=\hat{b}_{1}/\hat{b}_{2}$.
Because the costs $\hat{b}_{1}$ and $\hat{b}_{2}$ are positive, cross-multiplying turns the slope comparison into a comparison of profitabilities:
\begin{equation}\label{eqn:slope-profitability}
m_{I} > m_{B}
\;\Longleftrightarrow\;
\frac{\gamma_{1}}{\gamma_{2}} > \frac{\hat{b}_{1}}{\hat{b}_{2}}
\;\Longleftrightarrow\;
\frac{\gamma_{1}}{\hat{b}_{1}} > \frac{\gamma_{2}}{\hat{b}_{2}}
\;\Longleftrightarrow\;
\rho_{1} > \rho_{2}.
\end{equation}
Comparing the two slopes is therefore identical to ranking the substrates by profitability, and the optimum is the corner of the substrate with the larger $\rho_{i}$. 
Depending on the profitabilities $\rho_{i}$, the program either makes a distinct choice between sugars $i$ and $j$ (a corner solution) or, when the indifference curve runs exactly parallel to the budget line, spreads the synthesis budget across both (a degenerate set of optima).

When $m_{I}>m_{B}$, so that $\rho_{1}>\rho_{2}$, the optimum is the corner at point \texttt{b} (Fig.~\ref{fig:graphical-soln-ic-schematically}A), the highest utility the cell can reach on the budget line. Here the cell allocates all regulated synthesis to the more profitable enzyme 1: the target is $\hat{e}_{1}=\hat{b}_{1}^{-1}$, $\hat{e}_{2}=0$, and $u_{1}=\hat{b}_{1}\hat{e}_{1}=1$. Consumption of sugar 2 does not immediately stop, because its dynamic enzyme abundance and activity decay only gradually, but no further synthesis is committed to it.
When instead $m_{I}<m_{B}$, so that $\rho_{2}>\rho_{1}$, the optimum shifts to the corner at point \texttt{c} (Fig.~\ref{fig:graphical-soln-ic-schematically}B). Now all regulated synthesis goes to enzyme 2: $\hat{e}_{2}=\hat{b}_{2}^{-1}$, $\hat{e}_{1}=0$, and $u_{2}=\hat{b}_{2}\hat{e}_{2}=1$.

The remaining case, in which the two slopes are equal, is degenerate (Fig.~\ref{fig:graphical-soln-ic-schematically}C).
The profitabilities then coincide ($\rho_{1}=\rho_{2}$) and the indifference curve runs parallel to the budget constraint, so the cell solves:
\begin{align*}
\text{maximize} & \quad \sum_{i\in\mathcal{S}}\gamma_{i}\,\hat{e}_{i} \\
\text{subject to} & \quad \sum_{i\in\mathcal{S}}\hat{b}_{i}\,\hat{e}_{i} \leq 1 \\
\text{with} & \quad \rho_{i} = \rho\quad\text{for all}~i,
\end{align*}
where $\rho$ is their common value.
On the allocation simplex $u_{i}=\hat{b}_{i}\hat{e}_{i}$, rewriting the objective in these coordinates as $\gamma_{i}\hat{e}_{i} = (\gamma_{i}/\hat{b}_{i})\,u_{i} = \rho_{i}u_{i}$ gives $\sum_{i}\rho_{i}u_{i}$. With every $\rho_{i}=\rho$ this collapses to $\rho\sum_{i}u_{i} = \rho$, a constant, so every feasible allocation attains the optimum.
The optimal allocation is therefore not a point but the whole budget face, and demand becomes a set rather than a function.
Every allocation on the face, including the corners and every interior mixture, is optimal, and simultaneous substrate use (co-utilization) is among the optima.
Selecting a single allocation from this optimal face therefore requires an additional criterion, because a simplex solver returns a corner (a basic optimal solution) rather than an interior mixture even when the interior is equally optimal. A separate tie-breaking or regularization rule is needed to pick out the interior co-utilizing allocation; the matching law we derive next supplies one, its interior allocation lying on the optimal set and serving there as a selection rule rather than as a competing law.

\subsection*{Relationship to the classical matching law.}
The classical matching law is the same kind of consumer choice problem solved for a different utility, and writing both on the allocation simplex $\left\{u\geq{0},~\sum_{i}u_{i}=1\right\}$ with $u_{i}=\hat{b}_{i}e_{i}$ makes the contrast precise.
The linear program maximizes the linear (perfect-substitutes) utility $\sum_{i}\rho_{i}u_{i}$ in the profitabilities $\rho_{i}$. The maximum of a linear function over a simplex is attained at a vertex, so the solution is the single corner $i^{\star}=\arg\max_{i}\,\rho_{i}$.
The matching law $u_{i}=r_{i}/\sum_{k}r_{k}$ is, by contrast, the optimum of a Cobb--Douglas utility, where the $r_{i}$ are the substrate returns of Eqn.~\eqref{eqn:monod-like-rate}.
A Cobb--Douglas utility $W$ is a product of the allocations raised to fixed preference weights. Here the weight on substrate $i$ is its return $r_{i}$, so the cell values an allocation $u=\left(u_{1},\dotsc,u_{n}\right)$ through the expression:
\begin{equation}\label{eqn:cobb-douglas-utility}
W(u) = \prod_{i\in\mathcal{S}} u_{i}^{\,r_{i}},\qquad \ln W(u) = \sum_{i\in\mathcal{S}} r_{i}\ln u_{i}.
\end{equation}
Maximizing $W$ and its logarithm $\ln W$ return the same allocation, so we work with the additively separable log form.
The cell maximizes this utility while spending its entire synthesis budget, the simplex equality $\sum_{i}u_{i}=1$, so it solves:
\begin{align*}
\text{maximize} & \quad \ln W(u) = \sum_{i\in\mathcal{S}} r_{i}\ln u_{i} \\
\text{subject to} & \quad \sum_{i\in\mathcal{S}} u_{i} = 1 \\
\text{and} & \quad u_{i} \geq 0\qquad i=1,2,\dots,n.
\end{align*}
We solve this equality-constrained problem with a Lagrange multiplier.
Adjoining the budget equality $\sum_{i}u_{i}=1$ to the objective with a multiplier $\eta$ gives the Lagrangian:
\begin{equation}\label{eqn:matching-cobb-douglas}
\mathcal{L}(u,\eta) = \sum_{i\in\mathcal{S}} r_{i}\ln u_{i} - \eta\left(\sum_{i\in\mathcal{S}} u_{i} - 1\right),
\end{equation}
in which $\eta$ enforces the budget constraint. Assuming every return is strictly positive ($r_{i}>0$), the log utility is strictly concave and diverges to $-\infty$ as any $u_{i}\rightarrow 0$, so its maximizer over the simplex is interior and unique. When a return vanishes the corresponding allocation moves to the boundary, and when every return is zero the objective is constant and the cell commits no synthesis.
Setting the partial derivatives of $\mathcal{L}$ to zero gives the first-order (stationary) condition:
\begin{equation}\label{eqn:matching-foc}
\frac{\partial\mathcal{L}}{\partial u_{i}} = \frac{r_{i}}{u_{i}} - \eta = 0 \quad\Longrightarrow\quad u_{i} = \frac{r_{i}}{\eta}\qquad\forall i\in\mathcal{S}.
\end{equation}
Imposing the budget constraint then fixes the multiplier.
Summing $u_{i}=r_{i}/\eta$ over $i$ and setting the total to one gives $\sum_{i}r_{i}/\eta=1$, hence $\eta=\sum_{k}r_{k}$.
Back-substitution recovers the matching law:
\begin{equation}\label{eqn:matching-law}
u_{i} = \frac{r_{i}}{\sum_{k\in\mathcal{S}} r_{k}}\qquad\forall i\in\mathcal{S}.
\end{equation}
The cell allocates synthesis so that the marginal utility of allocation, $r_{i}/u_{i}$, is equalized across every substrate. That common value is the multiplier $\eta=\sum_{k}r_{k}$, the total return.
Each enzyme receives a share of the budget in proportion to its return $r_{i}$, an interior allocation.
This is a nonlinear (concave) program, not a linear program.

These two optima differ structurally.
The linear utility is finite at every vertex, so its maximum can sit at a corner, while the concave log utility attains its maximum in the interior.
The linear program thus commits the entire synthesis budget to one substrate, whereas the matching law spreads it across all competing substrates with positive return.
Each law is a limiting form of the constant-elasticity-of-substitution (CES) utility, the linear program at the perfect-substitutes limit and the matching law at the diversifying limit, as we show in Appendix~\ref{sec:ces}. The two limits carry different state-dependent weights, $\rho_{i}=\gamma_{i}/\hat{b}_{i}$ for the linear program and $r_{i}$ for the matching law, so they are not two members of one fixed-weight, one-parameter CES family. Both laws can instead be represented as limiting CES forms, and the elasticity of substitution is the axis along which they differ.
Young and Ramkrishna~\cite{Young:2007aa} also generalized the cybernetic laws, but along a different axis. Their generalized matching and proportional laws vary the return-on-investment structure within the interior allocation, whereas the elasticity-of-substitution family here varies substitutability, spanning the corner and interior regimes as its two limits.

\subsection*{Estimation of the expression cost coefficients.}
The corner solutions represent limiting cases that can be used to estimate values for the expression cost coefficients $\hat{b}_{i}$.
If the cell allocates all regulated synthesis to enzyme $i$ (or is presented with only sugar $i$), the intracellular environment approaches a steady state, approximated by balanced growth, and the dynamic abundance $e^{(i)}$ approaches the target $\hat{e}_{i}$.
The resource constraint then tells us that the target approaches the inverse of its cost coefficient, $\hat{b}_{i}\hat{e}_{i} \rightarrow 1$: the entire pseudo enzyme budget goes to enzyme $i$.
We can then eliminate the steady-state abundance of pseudo enzyme $i$ from the enzyme material balance equation, Eqn.~\eqref{eqn:enzyme-mat-bal}.
This gives an expression for the expression cost coefficient $\hat{b}_{i}$:
\begin{equation}\label{eqn:cost-coeff-approx}
\hat{b}_{i}^{-1} \approx \frac{1}{\delta_{i}}\cdot{\left(\frac{1}{\tau_{i}}+\lambda_{i}\right)}
\end{equation}
where we make use of the property that the cybernetic control variable $u_{i} \rightarrow 1$ in this condition, and define $\delta_{i} \equiv \left(\mu_{i}^{\text{max}} + \beta_{i}\right)$.
Eqn.~\eqref{eqn:cost-coeff-approx} rests on several assumptions. It takes the intracellular environment to be at steady state (balanced growth); it uses $u_{i}\rightarrow 1$; it replaces the realized dilution rate $\mu$ in the enzyme balance by the single-substrate maximum $\mu_{i}^{\text{max}}$; and it assumes substrate-saturated growth, so the Monod factor is near one. The profitability ranking $\rho_{i}=\gamma_{i}/\hat{b}_{i}$ inherits these assumptions; its sensitivity to the underlying kinetic and cost parameters is quantified by the robustness sweep reported in the Results.
However, it is a helpful approximation because it allows us to compute a characteristic value of the expression cost coefficients $\hat{b}_{i}$ from the model parameters $\tau_{i}$, $\lambda_{i}$, $\mu_{i}^{\text{max}}$, and $\beta_{i}$, which can be estimated from literature values or experimental data. Further, it gives us some insight into how the various model parameters influence the cost of gene expression.
For example, $\hat{b}_{i}$ is proportional to the characteristic time constant $\tau_{i}$: the longer it takes to express the pseudo enzyme $e_{i}(R)$, the more resource $R$ is consumed, and the higher $\hat{b}_{i}$ becomes.
Likewise, $\hat{b}_{i}$ is directly proportional to the dilution and degradation term $\delta_{i}$: the more unstable the pseudo enzyme $e_{i}(R)$, the more resource $R$ is consumed, and the higher $\hat{b}_{i}$ becomes.
Finally, $\hat{b}_{i}$ is inversely proportional to the rate of unregulated pseudo enzyme synthesis $\lambda_{i}$: as $\lambda_{i}$ increases, $\hat{b}_{i}$ decreases.
While this may seem counter-intuitive, it reflects the special case where the pseudo enzyme $e_{i}(R)$ is produced by a process not represented in the cybernetic model, such as a leaky promoter.
The unit cost of regulated production then decreases because part of the enzyme is supplied at no modeled resource cost.

The constitutive expression rate $\lambda_{i}$ governs how close a substrate pair sits to the degenerate face.
From Eqn.~\eqref{eqn:cost-coeff-approx}, increasing $\lambda_{i}$ reduces $\hat{b}_{i}$, which raises the profitability ratio $\gamma_{i}/\hat{b}_{i}$ for substrate $i$ toward that of its competitor.
When the two ratios equalize, the pair reaches the degenerate face, where every allocation on the budget line is optimal and co-utilization is one of the optima.

\subsection*{Parameter estimation and calibration.}
The parameters came from two sources (Table~\ref{tbl:parameters}). The maximum growth rates $\mu_{i}^{\text{max}}$, the saturation constants $K_{i}$, the degradation rates $\beta_{i}$, and the glucose, lactose, and fructose yields came from the single-substrate characterization of Kompala et al.~\cite{Kompala:1986aa} and were held fixed. The model-level quantities, the common expression time constant $\tau$, the enzyme scaling levels $e_{i}^{\text{max}}$, and the cell death rate, were not measured per substrate. We set them by hand so that the integrated model matched the observed mixed-substrate cellmass trajectories. The diauxic and triauxic growth curves of Fig.~\ref{fig:composite-diauxic} and Fig.~\ref{fig:composite-triauxic} were therefore used to calibrate these values, not held out as an independent test. We calibrated the xylose yield $Y_{2}$ and scaling level $e_{2}^{\text{max}}$ separately for the two mixtures because the diauxic and triauxic data came from different batch experiments in Kompala et al.~\cite{Kompala:1986aa}. We set the constitutive synthesis rates $\lambda_{i}$ to zero for the regulated enzymes. The one case with a nonzero $\lambda$, the fructose enzyme in the glucose--fructose analysis, demonstrated the co-utilization mechanism and was not a fit to fructose growth data. The calibration was manual and targeted the cellmass trajectory, so the model-level values were characteristic rather than uniquely identified, and we did not perform a formal uncertainty or identifiability analysis.

The calibration did not set the substrate preference order. That order came from the profitability ranking $\gamma_{i}/\hat{b}_{i}$ (Table~\ref{tbl:profitability}), which used only the single-substrate kinetic parameters and the cost coefficients of Eqn.~\eqref{eqn:cost-coeff-approx}. The robustness sweep (Fig.~\ref{fig:profitability-robustness}) showed that this ranking held under $\pm 25\%$ perturbation of the parameters that entered it. The order of substrate use was therefore a prediction of the LP geometry, while the calibrated model-level parameters set the timing and shape of the trajectory.

\subsection*{Computational methods.}
We integrated the governing ODEs forward in time with a forward Euler scheme at step size $h = 0.01$ h. At each step we solved the resource-allocation LP with \texttt{GLPK}~v1.2.1~\cite{GLPK} at its default feasibility and optimality tolerances, obtaining the cybernetic $u$-variables.
A substrate whose concentration fell below the depletion threshold $s_{\text{dep}} = 10^{-3}$ g/L was treated as exhausted, with its return $\gamma_{i}$ set to zero (Eqn.~\eqref{eqn:gamma-def}) so that it left the allocation problem; because substrate concentrations are monotonically non-increasing during batch growth, this gate acts as a latch. The same rule is applied when the controls are recorded for the figures, so the plotted $u$-variables are the controls used during integration.
We verified step-size convergence at $h = 0.01$, $0.005$, and $0.0025$ h. Halving the step shifted the predicted glucose-depletion time, the diauxic transition of Fig.~\ref{fig:composite-diauxic}, from 4.14 h to 4.1325 h, a change of under 0.2\%, so the reported trajectories were step-size independent.
Goodness of fit was quantified by the root-mean-square error (RMSE) and coefficient of determination ($R^{2}$), both computed on $\log_{10}$ cellmass; model predictions were linearly interpolated to the observation times, and the experimental cellmass curves were digitized from the published figures of Kompala et al.~\cite{Kompala:1986aa}.
All simulations were performed in Julia~1.12.6~\cite{Bezanson:2017aa} using JuMP~v1.30.1~\cite{Lubin:2023aa} and Plots~v1.41.6.

%% file: sections/results.tex
\begin{figure}[h!]
\begin{center}
\includegraphics[height=0.68\textheight]{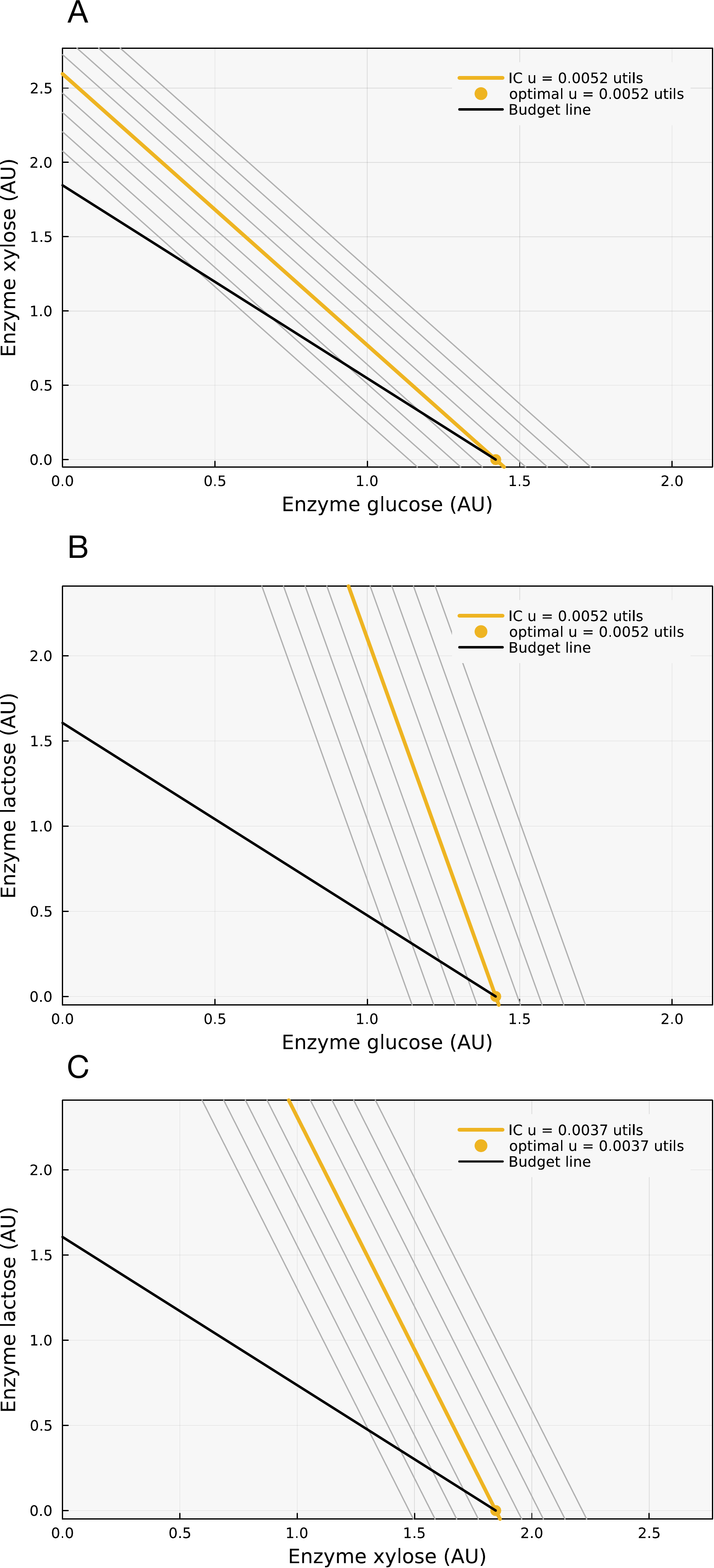}
\end{center}
\caption{Steady-state cybernetic control problem solutions for the three binary sugar choices among glucose, xylose, and lactose.
The budget line (black line) represents the resource constraint,
indifference curves (gray lines) are lines of constant utility, where the yellow indifference curve represents the highest possible utility the cell can achieve given the resource constraint.
Indifference curves were computed at utility levels $U \in \left\{\theta\cdot{u^{\star}}\right\}$, where $u^{\star}$ is the optimal utility computed by solving the resource allocation problem, and $\theta\in\left[0.8,1.2\right]$.
In all cases, $\lambda_{i} = 0$ and the sugar concentrations were $s_{1} = 0.33$ g/L (glucose), $s_{2} = 2.0$ g/L (xylose), and $s_{3} = 1.5$ g/L (lactose) (reproduced from Kompala et al.~\cite{Kompala:1986aa}).
These snapshot concentrations differ from the experimental initial conditions used in the dynamic simulations (glucose 0.5 g/L; xylose 2.5 g/L or 1.5 g/L; lactose 5.0 g/L). The predicted preference order glucose $>$ xylose $>$ lactose holds at both the snapshot and the dynamic initial concentrations, as confirmed by the profitability ratios in Table~\ref{tbl:profitability}.
\textbf{A}: Budget line and indifference curves for the glucose (horizontal) and xylose (vertical) system. Glucose is predicted to be the preferred sugar.
\textbf{B}: Budget line and indifference curves for the glucose (horizontal) and lactose (vertical) system. Glucose is predicted to be the preferred sugar.
\textbf{C}: Budget line and indifference curves for the xylose (horizontal) and lactose (vertical) system. Xylose is predicted to be the preferred sugar.}\label{fig:graphical-soln-ic-computed}
\end{figure}
\begin{figure}[h!]
\begin{center}
\includegraphics[width=0.95\textwidth]{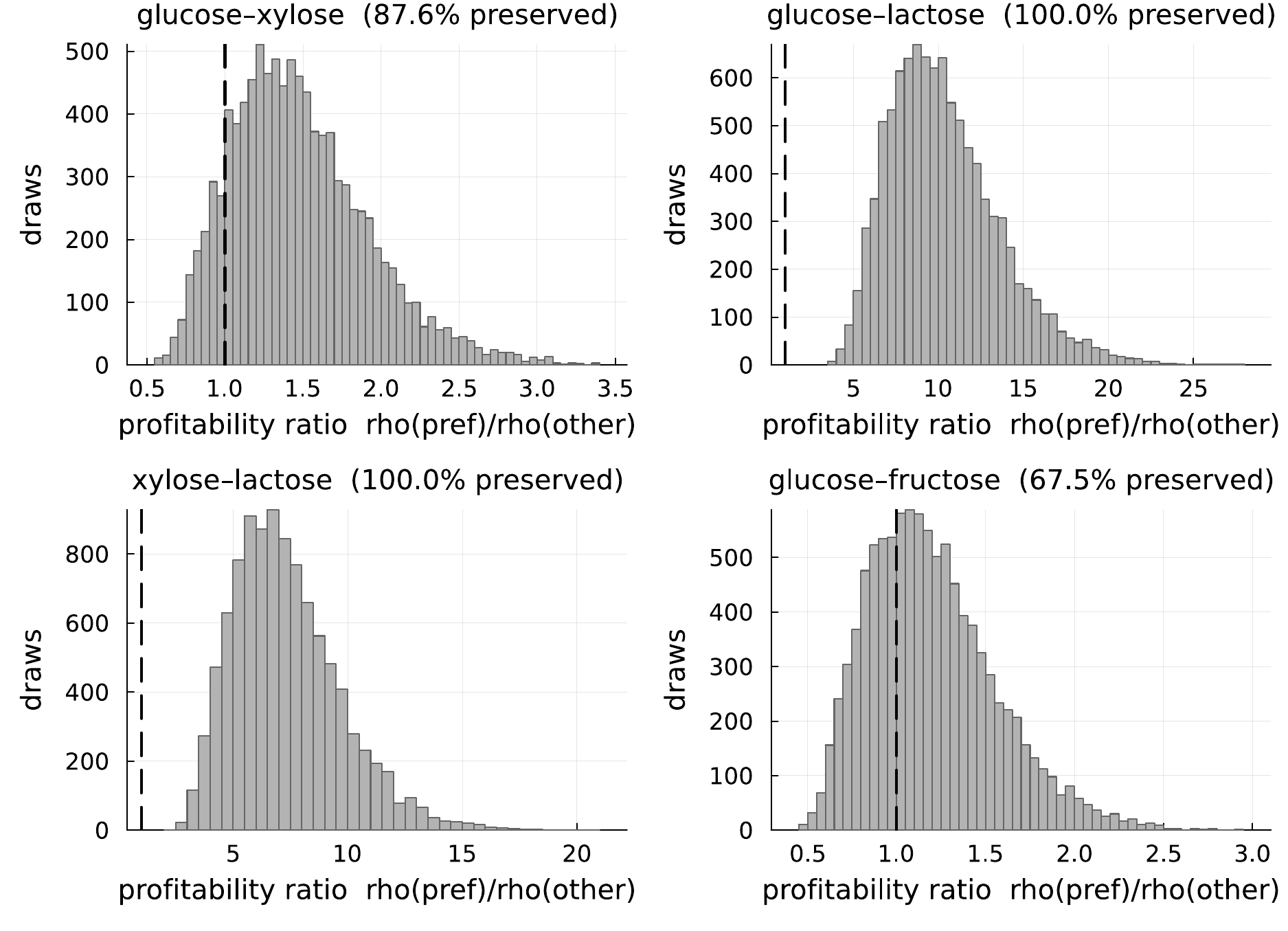}
\end{center}
\caption{Robustness of the predicted preference order to parameter uncertainty.
Distribution of the profitability ratio $\gamma_{i}/\hat{b}_{i}$ of the preferred over the competing substrate across $10{,}000$ parameter sets, each drawn by perturbing the maximum growth rate $\mu_{i}^{\text{max}}$, saturation constant $K_{i}$, expression time constant $\tau_{i}$, degradation rate $\beta_{i}$, and enzyme scaling level $e_{i}^{\text{max}}$ independently by $\pm 25\%$ about the values of Table~\ref{tbl:parameters}. The dashed vertical line marks the flip boundary (ratio $=1$); mass to its right indicates the nominal preference order is preserved. The three diauxic pairs (glucose--xylose, glucose--lactose, xylose--lactose) lie predominantly to the right of the boundary, while glucose--fructose straddles it, consistent with its near-degenerate profitability.
}\label{fig:profitability-robustness}
\end{figure}

We solved the LP for each binary sugar pair and recovered a strict order of substrate preference (Fig.~\ref{fig:graphical-soln-ic-computed}). We computed the budget coefficients $\hat{b}_{i}$ and utility coefficients $\gamma_{i}$ from the model parameters (Table~\ref{tbl:parameters}) and initial conditions taken from Kompala et al.~\cite{Kompala:1986aa}. Glucose was the preferred substrate over both xylose (Fig.~\ref{fig:graphical-soln-ic-computed}A) and lactose (Fig.~\ref{fig:graphical-soln-ic-computed}B), and xylose was preferred over lactose (Fig.~\ref{fig:graphical-soln-ic-computed}C). The LP returned the correct corner solution in every case, consistent with the experimental consumption order reported by Kompala et al.~\cite{Kompala:1986aa}. In each binary pair, the slope of the indifference curves exceeded the slope of the budget line, placing the optimum at the corner corresponding to the more profitable substrate and allocating the entire enzyme synthesis budget to that substrate alone. The profitability ratios $\gamma_i/\hat{b}_i$ made this ordering explicit, with the preferred substrate given by the argmax in every pair (Table~\ref{tbl:profitability}). This geometric result established the predicted order of substrate preference. That ordering was robust to parameter uncertainty (Fig.~\ref{fig:profitability-robustness}). We drew $10{,}000$ parameter sets (fixed random seed 42) by perturbing the five kinetic and cost parameters that enter the profitability ratio, the maximum growth rate $\mu_{i}^{\text{max}}$, saturation constant $K_{i}$, expression time constant $\tau_{i}$, degradation rate $\beta_{i}$, and enzyme scaling level $e_{i}^{\text{max}}$, each by an independent multiplicative factor drawn uniformly on $\pm 25\%$ about the values of Table~\ref{tbl:parameters}, and recomputed the profitability ratio $\gamma_{i}/\hat{b}_{i}$ for each binary pair. The draws were independent across parameters; parameter correlations and structural uncertainty were not considered. The preferred substrate, the argmax of the profitability ratio, was unchanged in 87.6\% of draws for glucose--xylose, 100\% for glucose--lactose, and 100\% for xylose--lactose, placing the three diauxic pairs away from the degeneracy boundary. The glucose--fructose pair, by contrast, straddled the boundary, with glucose preferred in 67.5\% of draws, consistent with its near-degenerate profitability ratios (Table~\ref{tbl:profitability}) and the co-utilization behavior analyzed below. We then tested whether the LP, solved at each integration step, allowed the ODE system to reproduce the observed diauxic and triauxic batch growth cellmass trajectories.

\begin{figure}[h!]
\begin{center}
\includegraphics[width=0.99\textwidth]{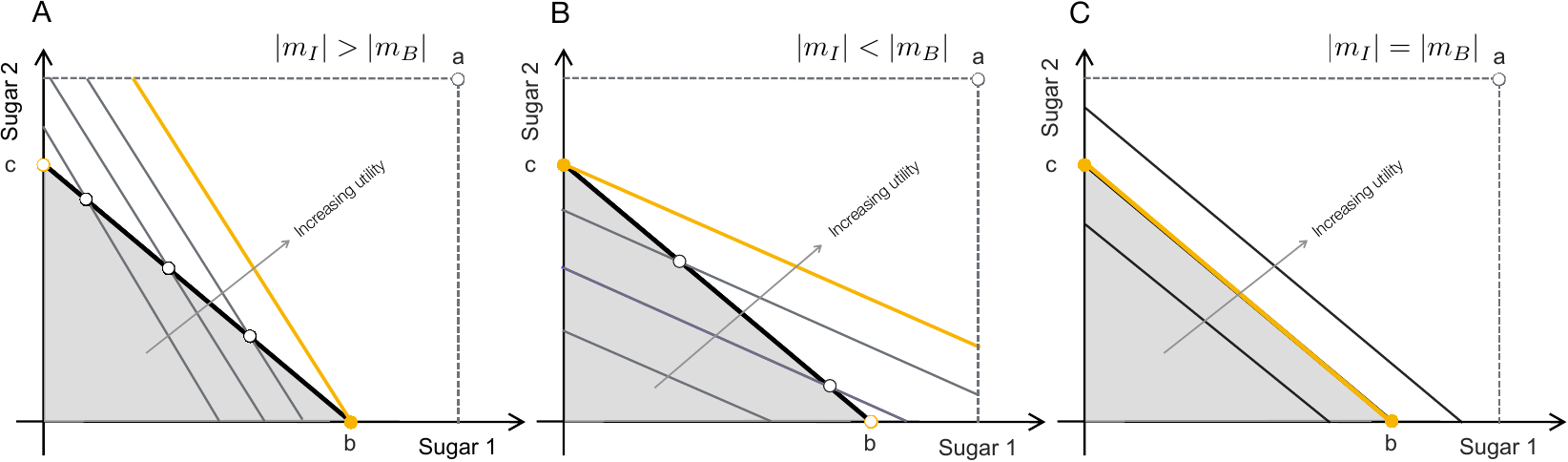}
\end{center}
\caption{Schematic of the solution regimes for the cybernetic control problem for a two-sugar system.
The horizontal and vertical axes represent the target abundances $\hat{e}_{1}$ and $\hat{e}_{2}$ of the pseudo enzymes that consume sugars 1 and 2.
Gray and yellow lines represent the cell's indifference curves (lines of constant utility).
The black line represents the resource constraint, and the gray region is the feasible set, the allocations for which the resource constraint holds as an inequality.
Point \texttt{a} represents the highest possible utility for the cell. However, the cell cannot achieve this utility because it does not have enough resources to express both pseudo enzymes $e_{1}(R)$ and $e_{2}(R)$
at this level.
Point \texttt{b} represents allocating all regulated synthesis to enzyme 1, where the target $\hat{e}_{1} = \hat{b}_{1}^{-1}$ and $\hat{e}_{2}=0$.
Point \texttt{c} represents allocating all regulated synthesis to enzyme 2, where $\hat{e}_{2} = \hat{b}_{2}^{-1}$ and $\hat{e}_{1}=0$.
\textbf{A}: The slope of the indifference curves $m_{I}$ is greater than the slope of the resource constraint $m_{B}$; thus, the cell allocates all synthesis to enzyme 1 (a corner solution at point \texttt{b}).
\textbf{B}: The slope of the indifference curves $m_{I}$ is less than the slope of the resource constraint $m_{B}$; thus, the cell allocates all synthesis to enzyme 2 (a corner solution at point \texttt{c}).
\textbf{C}: The slope of the indifference curves $m_{I}$ equals the slope of the resource constraint $m_{B}$; the cell can split its synthesis budget across both enzymes, and every allocation along the resource constraint is optimal (the degenerate case).
}\label{fig:graphical-soln-ic-schematically}
\end{figure}

\begin{table}
	\centering
		\caption{Parameters for the cybernetic model of \emph{K.~oxytoca} batch growth on mixed substrates.
		Glucose parameters are identical across both simulation configurations.
		Xylose parameters differ between the diauxic (glucose--xylose, Fig.~\ref{fig:composite-diauxic}) and triauxic (glucose--xylose--lactose, Fig.~\ref{fig:composite-triauxic}) experiments; both values are given where they differ (diauxic / triauxic).
		Lactose parameters are taken from the triauxic configuration only.
		Fructose parameters are used only in the glucose--fructose co-utilization analysis (Fig.~\ref{fig:glucose-fructose}, Fig.~\ref{fig:v-ablation}).
		Parameters marked ``this study'' are model-level quantities that were not measured per substrate: the common expression time constant $\tau$ and the enzyme scaling levels $e_{i}^{\text{max}}$ were set in this work, and the xylose yield $Y_{2}$ and scaling level $e_{2}^{\text{max}}$ were re-calibrated for each mixed-substrate experiment (diauxic/triauxic values shown). These model-level values were tuned by hand so that the simulated cellmass matched the observed mixed-substrate trajectories, which were therefore used for calibration rather than held out. The kinetic parameters ($\mu_{i}^{\text{max}}$, $K_{i}$, $\beta_{i}$, and the glucose, lactose, and fructose yields) are from the single-substrate characterization of Kompala et al.~\cite{Kompala:1986aa}.}
		\begin{tabular}{lcccc} \toprule
			\textbf{Description} & \textbf{Parameter} & \textbf{Value} & \textbf{Units} & \textbf{Reference} \\ \toprule
			Glucose maximum growth rate & $\mu_{1}^{\text{max}}$ & 1.08 & 1/h & \cite{Kompala:1986aa} \\
			Glucose saturation constant & $K_{1}$ & 0.01 & g/L & \cite{Kompala:1986aa} \\
			Glucose yield coefficient & $Y_{1}$ & 0.52 & gDW/g & \cite{Kompala:1986aa} \\
			Glucose expression time constant & $\tau_{1}$ & 0.623 & h & \text{this study} \\
			Glucose enzyme degradation rate & $\beta_{1}$ & 0.05 & 1/h & \cite{Kompala:1986aa} \\
			Glucose enzyme scaling level & $e_{1}^{max}$ & 1.42 & -- & \text{this study} \\
			Glucose unregulated enzyme synthesis rate & $\lambda_{1}$ & 0.0 & 1/h & \text{this study} \\

			& & & & \\
			Xylose maximum growth rate & $\mu_{2}^{\text{max}}$ & 0.82 & 1/h & \cite{Kompala:1986aa} \\
			Xylose saturation constant & $K_{2}$ & 0.20 & g/L & \cite{Kompala:1986aa} \\
			Xylose yield coefficient$^{\dagger}$ & $Y_{2}$ & 0.58 / 0.50 & gDW/g & \text{this study} \\
			Xylose expression time constant & $\tau_{2}$ & 0.623 & h & \text{this study} \\
			Xylose enzyme degradation rate & $\beta_{2}$ & 0.05 & 1/h & \cite{Kompala:1986aa} \\
			Xylose enzyme scaling level$^{\dagger}$ & $e_{2}^{max}$ & 1.85 / 3.46 & -- & \text{this study} \\
			Xylose unregulated enzyme synthesis rate & $\lambda_{2}$ & 0.0 & 1/h & \text{this study} \\

			& & & & \\
			Lactose maximum growth rate & $\mu_{3}^{\text{max}}$ & 0.95 & 1/h & \cite{Kompala:1986aa} \\
			Lactose saturation constant & $K_{3}$ & 4.5 & g/L & \cite{Kompala:1986aa} \\
			Lactose yield coefficient & $Y_{3}$ & 0.45 & gDW/g & \cite{Kompala:1986aa} \\
			Lactose expression time constant & $\tau_{3}$ & 0.623 & h & \text{this study} \\
			Lactose enzyme degradation rate & $\beta_{3}$ & 0.05 & 1/h & \cite{Kompala:1986aa} \\
			Lactose enzyme scaling level & $e_{3}^{max}$ & 3.61 & -- & \text{this study} \\
			Lactose unregulated enzyme synthesis rate & $\lambda_{3}$ & 0.0 & 1/h & \text{this study} \\

			& & & & \\
			Fructose maximum growth rate & $\mu_{4}^{\text{max}}$ & 0.94 & 1/h & \cite{Kompala:1986aa} \\
			Fructose saturation constant & $K_{4}$ & 0.01 & g/L & \cite{Kompala:1986aa} \\
			Fructose yield coefficient & $Y_{4}$ & 0.52 & gDW/g & \cite{Kompala:1986aa} \\
			Fructose expression time constant & $\tau_{4}$ & 0.623 & h & \text{this study} \\
			Fructose enzyme degradation rate & $\beta_{4}$ & 0.05 & 1/h & \cite{Kompala:1986aa} \\
			Fructose enzyme scaling level & $e_{4}^{max}$ & 1.62 & -- & \text{this study} \\
			Fructose unregulated enzyme synthesis rate & $\lambda_{4}$ & 0.0 & 1/h & \text{this study} \\

			& & & & \\
			Cell death rate constant & $k_{d}^{c}$ & 0.022 & 1/h & \text{this study} \\
			\bottomrule
			\multicolumn{5}{l}{$^{\dagger}$~Diauxic (glucose--xylose) value / triauxic (glucose--xylose--lactose) value.}
		\end{tabular}
		\label{tbl:parameters}
\end{table}

\begin{table}
	\centering
	\caption{Initial conditions used in the dynamic simulations.
	Enzyme levels $e_i(0)$ are expressed as fractions of their respective reference scaling levels $e_i^{max}$.
	Under $\pm 50\%$ perturbation of the initial xylose enzyme level $e_2(0)$, the diauxic lag onset was unchanged to the reported precision (4.24 h, 4.24 h, 4.24 h; one integration step is $0.01$ h), indicating that lag timing is set by corner-switching and basal synthesis kinetics rather than by the initial enzyme level.}
	\label{tbl:initial-conditions}
	\begin{tabular}{lccc} \toprule
		\textbf{Quantity} & \textbf{Glucose--xylose} & \textbf{Glucose--xylose--lactose} & \textbf{Units} \\ \toprule
		$s_{\text{glc}}(0)$ & 0.5 & 0.5 & g/L \\
		$s_{\text{xyl}}(0)$ & 2.5 & 1.5 & g/L \\
		$s_{\text{lac}}(0)$ & -- & 5.0 & g/L \\
		$e_{\text{glc}}(0)$ & $0.90\,e_{\text{glc}}^{max}$ & $0.90\,e_{\text{glc}}^{max}$ & -- \\
		$e_{\text{xyl}}(0)$ & $0.18\,e_{\text{xyl}}^{max}$ & $0.17\,e_{\text{xyl}}^{max}$ & -- \\
		$e_{\text{lac}}(0)$ & -- & $0.20\,e_{\text{lac}}^{max}$ & -- \\
		$c(0)$ & $4.0\times10^{-3}$ & $2.1\times10^{-3}$ & gDW/L \\
		\bottomrule
	\end{tabular}
\end{table}

\begin{table}
	\centering
	\caption{Profitability ratios $\gamma_i/\hat{b}_i$ for each binary sugar pair.
	The preferred substrate is $i^{\star} = \arg\max_i\,\gamma_i/\hat{b}_i$, the closed-form argmax from the LP corner analysis.
	Values are computed from the parameters in Table~\ref{tbl:parameters} at the static-analysis concentrations of Fig.~\ref{fig:graphical-soln-ic-computed} (glucose 0.33, xylose 2.0, lactose 1.5, fructose 0.33 g/L) using Eqn.~\eqref{eqn:gamma-def} and Eqn.~\eqref{eqn:cost-coeff-approx}. Because each $\gamma_i$ depends on its substrate concentration, these ratios are specific to the stated concentrations; the preference ordering also holds at the dynamic initial concentrations, whereas the glucose--fructose degeneracy point is state-specific, not concentration-independent.}
	\label{tbl:profitability}
	\begin{tabular}{llcccc} \toprule
		\textbf{Pair} & \textbf{Sugar} & $\gamma_i$ & $\hat{b}_i$ & $\gamma_i/\hat{b}_i$ & \textbf{Preferred} \\ \toprule
		glucose--xylose & glucose  & 0.73748 & 0.70354 & 1.04824 & glucose \\
		                & xylose   & 0.40379 & 0.54167 & 0.74545 & \\
		\midrule
		glucose--lactose & glucose & 0.73748 & 0.70354 & 1.04824 & glucose \\
		                 & lactose & 0.06586 & 0.62261 & 0.10578 & \\
		\midrule
		xylose--lactose & xylose  & 0.40379 & 0.54167 & 0.74545 & xylose \\
		                & lactose & 0.06586 & 0.62261 & 0.10578 & \\
		\midrule
		glucose--fructose & glucose  & 0.73748 & 0.70354 & 1.04824 & glucose \\
		                  & fructose & 0.56236 & 0.61638 & 0.91235 & \\
		\bottomrule
	\end{tabular}
\end{table}

\begin{figure}[h!]
\begin{center}
\includegraphics[width=0.95\textwidth]{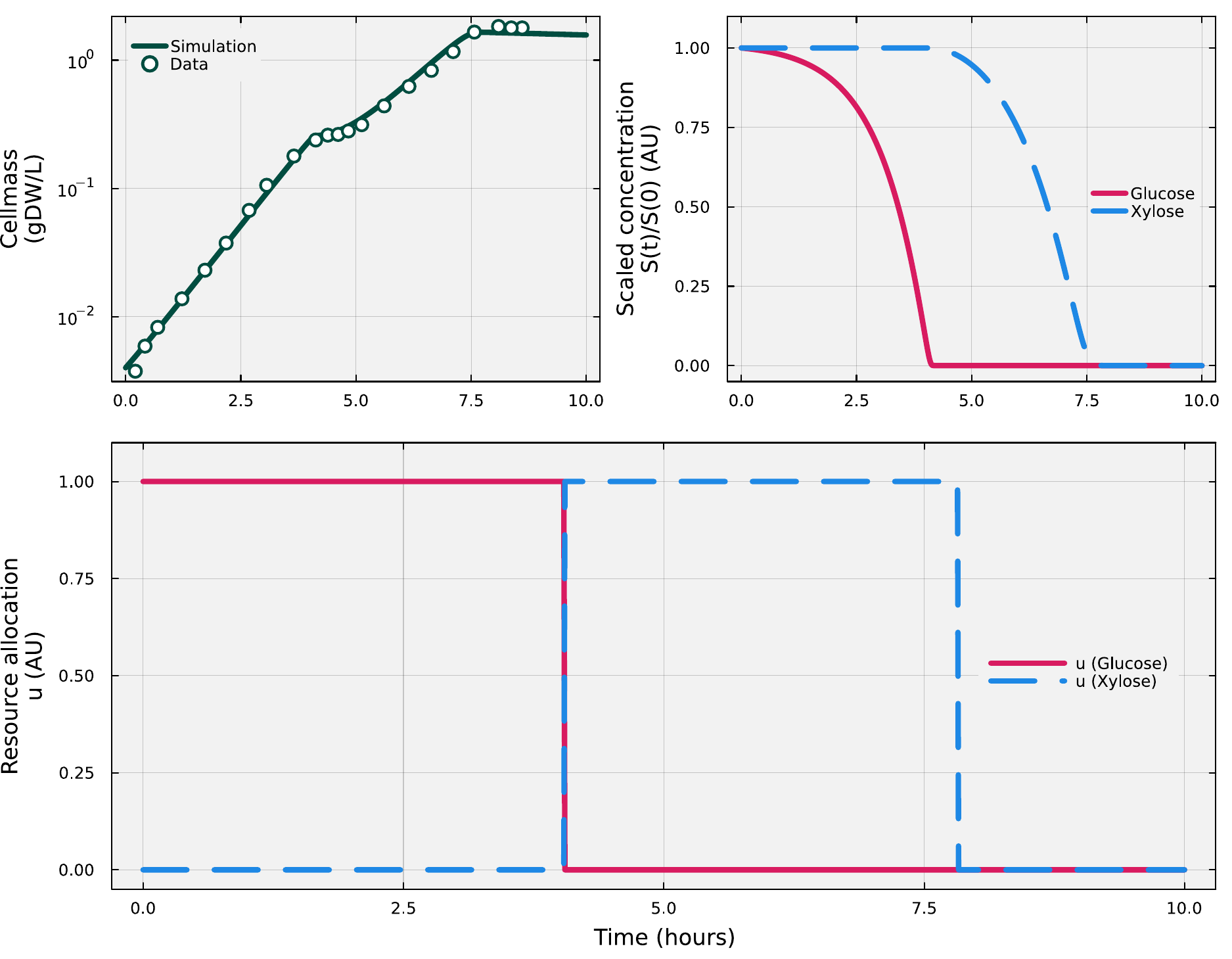}
\end{center}
\caption{LP-driven cybernetic model simulation of diauxic batch growth of \emph{Klebsiella oxytoca} on glucose and xylose.
\textbf{(A)} Cellmass concentration (log scale): simulation (line) vs.\ experimental data from Kompala et al.~\cite{Kompala:1986aa} (circles).
\textbf{(B)} Scaled substrate concentrations $S(t)/S(0)$: glucose (pink) and xylose (blue) are consumed sequentially.
\textbf{(C)} Cybernetic $u$-variable dynamics: the LP switches from the glucose corner ($u_{\text{glc}}=1$) to the xylose corner ($u_{\text{xyl}}=1$) upon glucose exhaustion, with no intermediate allocation.
The step-function switching shows the LP returning a corner solution at each time step; these are the controls recorded during integration.
}\label{fig:composite-diauxic}
\end{figure}

\begin{figure}[h!]
\begin{center}
\includegraphics[width=0.65\textwidth]{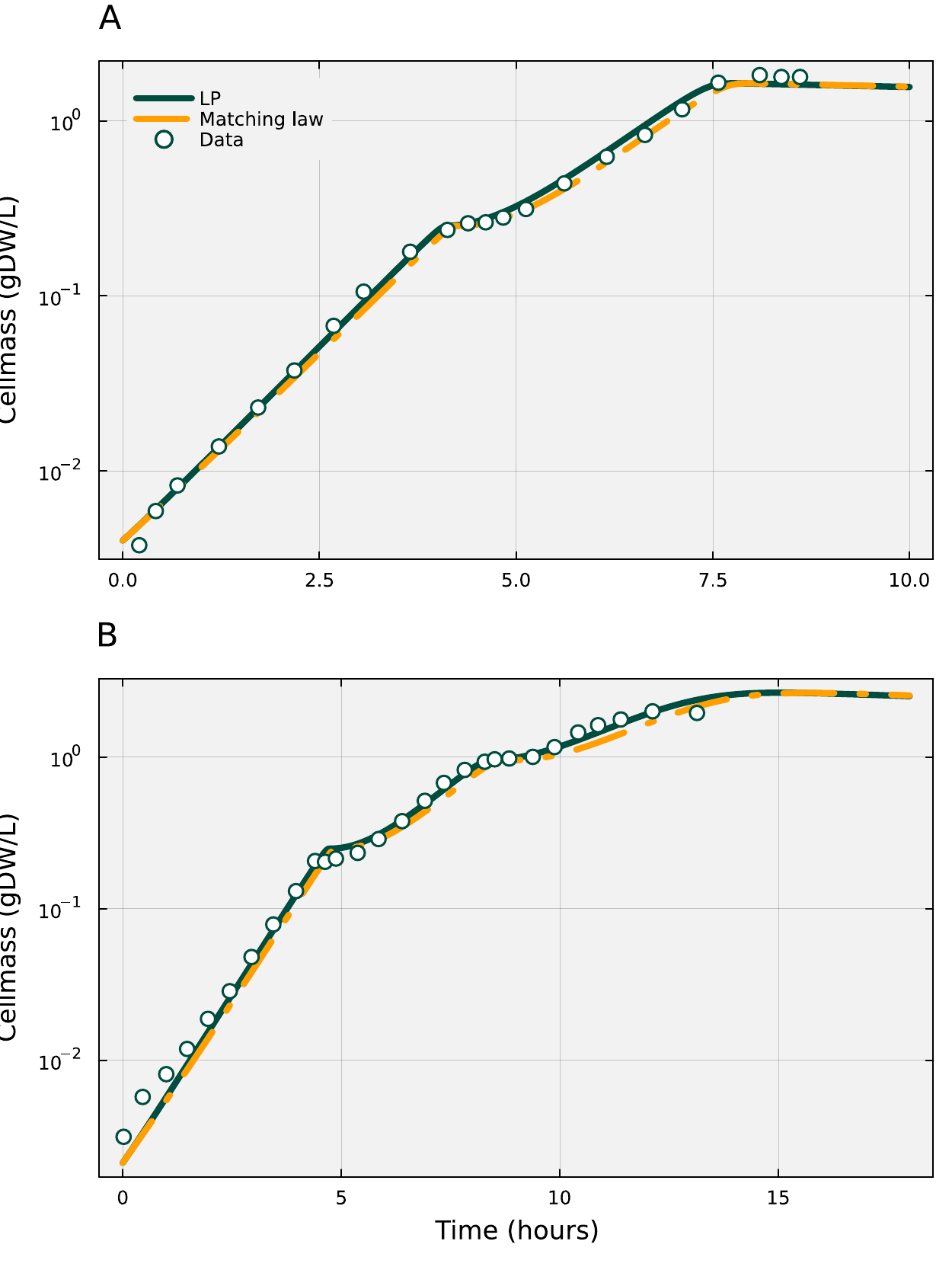}
\end{center}
\caption{Cellmass trajectories under the LP (solid) and the classical matching law (dash-dot) against the data of Kompala et al.~\cite{Kompala:1986aa} (circles), on a log scale.
\textbf{(A)} Glucose--xylose (diauxic). LP fit RMSE = 0.0421, $R^{2}$ = 0.9974; matching law RMSE = 0.0461, $R^{2}$ = 0.9969.
\textbf{(B)} Glucose--xylose--lactose (triauxic). LP fit RMSE = 0.0830, $R^{2}$ = 0.9900; matching law RMSE = 0.1123, $R^{2}$ = 0.9816.
Both control laws reproduced the cellmass curves; the LP was marginally better, with the differences concentrated at the diauxic transitions.}\label{fig:lp-vs-matching}
\end{figure}

We simulated diauxic batch growth of \emph{K.~oxytoca} on glucose (0.5 g/L) and xylose (2.5 g/L) (Table~\ref{tbl:initial-conditions}), and found that the LP-driven model quantitatively reproduced the observed diauxic growth using kinetic parameters fixed from single-substrate experiments together with the model-level parameters of Table~\ref{tbl:parameters} (Fig.~\ref{fig:composite-diauxic}). The simulated cellmass tracked the measured data through both exponential phases, including the characteristic growth pause near $t \approx 4.1$ h (Fig.~\ref{fig:composite-diauxic}A).
The simulated substrate profiles confirmed strict sequential depletion (glucose was exhausted near $t \approx 4.1$ h before xylose consumption accelerated) in correspondence with the corner solution predicted by the static LP analysis (Fig.~\ref{fig:composite-diauxic}B). The recorded $u$-variable time series showed that the LP held the glucose corner ($u_{\text{glc}} = 1$, $u_{\text{xyl}} = 0$) throughout the glucose phase and switched sharply to the xylose corner ($u_{\text{glc}} = 0$, $u_{\text{xyl}} = 1$) upon glucose exhaustion, with no intermediate allocation at any time step (Fig.~\ref{fig:composite-diauxic}C). These step functions indicated that the sequence of diauxic growth followed from the geometry of the LP rather than from an explicit regulatory rule. On a log cellmass scale the LP fit gave RMSE = 0.0421 and $R^{2}$ = 0.9974, comparable to the classical cybernetic matching law (RMSE = 0.0461, $R^{2}$ = 0.9969) on the same trajectory (Fig.~\ref{fig:lp-vs-matching}A). Thus, the two laws fit the cellmass curve similarly, as expected since the activity control was identical and the synthesis laws differed mainly during the brief transition. The near-coincidence indicated that the diauxic phenotype was a consequence of corner-switching rather than of the specific synthesis law.

\begin{figure}[h!]
\begin{center}
\includegraphics[width=0.95\textwidth]{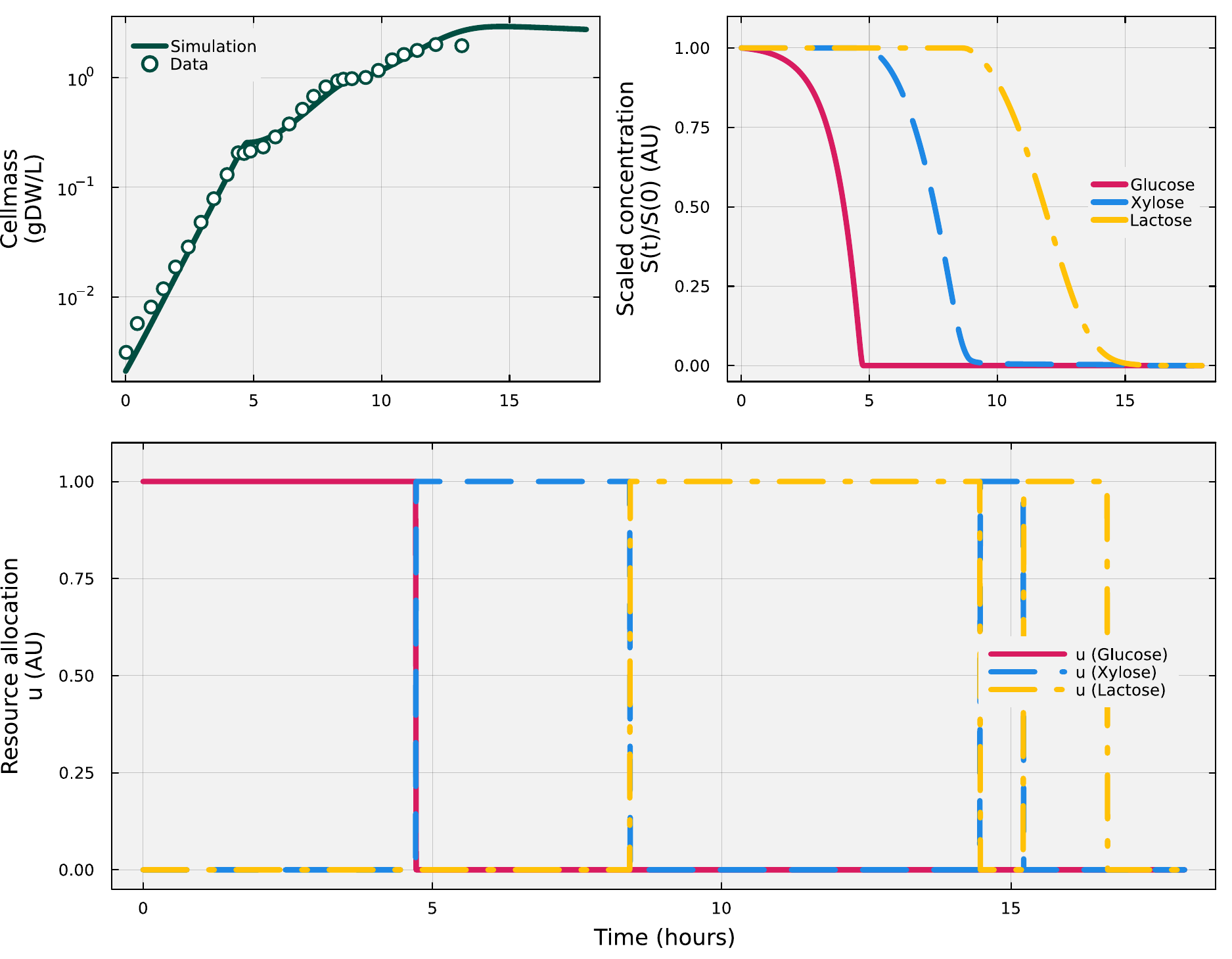}
\end{center}
\caption{LP-driven cybernetic model simulation of triauxic batch growth of \emph{Klebsiella oxytoca} on glucose, xylose, and lactose.
\textbf{(A)} Cellmass concentration (log scale): simulation (line) vs.\ experimental data from Kompala et al.~\cite{Kompala:1986aa} (circles).
\textbf{(B)} Scaled substrate concentrations $S(t)/S(0)$: glucose (pink), xylose (blue), and lactose (gold) are consumed sequentially.
\textbf{(C)} Cybernetic $u$-variable dynamics: the LP visits three corner solutions in sequence (glucose, xylose, lactose) with sharp transitions at each substrate exhaustion.
The order of LP corner visits matches the substrate preference order predicted by the static LP analysis (Fig.~\ref{fig:graphical-soln-ic-computed}).
}\label{fig:composite-triauxic}
\end{figure}

Having established that LP corner-switching reproduced diauxic growth, we next asked whether the same framework extended to three substrates (Fig.~\ref{fig:composite-triauxic}). We simulated triauxic batch growth of \emph{K.~oxytoca} on glucose (0.5 g/L), xylose (1.5 g/L), and lactose (5.0 g/L) and found that the LP-driven model reproduced the full three-phase cellmass growth curve, capturing both successive diauxic lags near $t \approx 5$ h and $t \approx 8$ h, using kinetic parameters fixed from single-substrate experiments together with the model-level parameters of Table~\ref{tbl:parameters} (Fig.~\ref{fig:composite-triauxic}A). The simulated substrate profiles showed strict sequential depletion in the order glucose $\rightarrow$ xylose $\rightarrow$ lactose, in agreement with the profitability ranking $\gamma_i/\hat{b}_i$ established by the static LP analysis (Fig.~\ref{fig:composite-triauxic}B). The $u$-variable time series showed three non-overlapping unit-height step functions (Fig.~\ref{fig:composite-triauxic}C), and the LP visited the glucose, xylose, and lactose corners in order, in correspondence with the static preference order. The two diauxic lags corresponded to the transient periods required to accumulate sufficient enzyme after each corner switch. On a log scale the LP fit gave RMSE = 0.0830 and $R^{2}$ = 0.9900, again comparable to the classical cybernetic matching law (RMSE = 0.1123, $R^{2}$ = 0.9816) on the same trajectory (Fig.~\ref{fig:lp-vs-matching}B).
The LP was marginally better, with the difference concentrated at the transition points where the synthesis laws diverged.
Taken together, the diauxic and triauxic results were consistent with a single LP whose profitability ranking predicted the order of substrate use from single-substrate kinetics, while a small set of model-level parameters, calibrated to each mixture, set the quantitative timing and dynamics of multi-substrate growth.

\begin{figure}[h!]
\begin{center}
\includegraphics[width=0.95\textwidth]{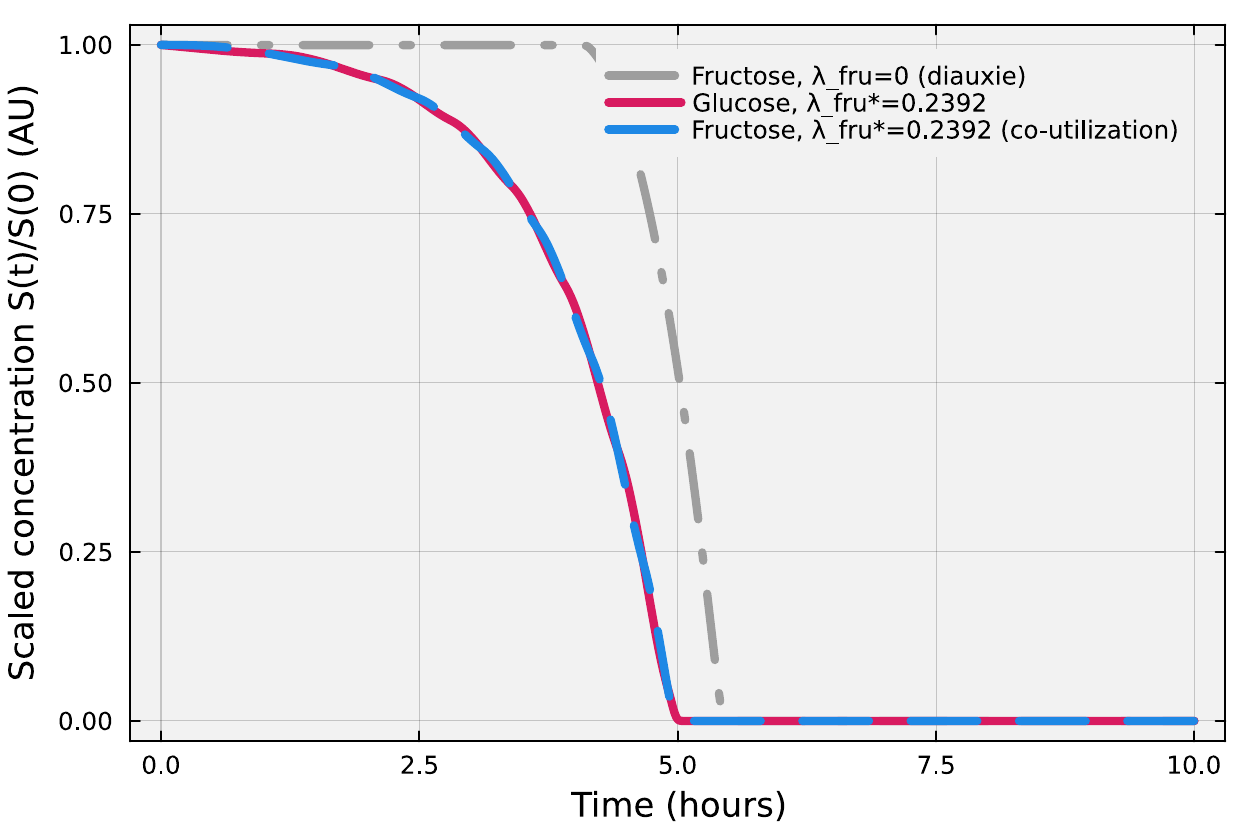}
\end{center}
\caption{Glucose--fructose dynamics under the linear program (scaled substrate concentrations). At $\lambda_{\text{fru}}=0$ the model predicts diauxie (fructose, dash-dot, untouched until glucose is gone). Raising the constitutive expression rate to the specified-state degeneracy value ($\lambda_{\text{fru}}^{\star}=0.2392$, computed at the initial concentrations) lowers the fructose proteome cost until the profitabilities are equal at that state, and glucose (solid) and fructose (dashed) are then consumed simultaneously. At this point the LP still returns corner solutions; the simultaneity reflects rapid corner alternation between the two equally profitable substrates together with the constitutive fructose expression, not an interior allocation. This rate far exceeds the near-zero value of the regulated enzymes, so it illustrates the mechanism rather than a fit to the glucose--fructose data.}\label{fig:glucose-fructose}
\end{figure}

Not all sugar pairs were strictly ordered. The glucose--fructose pair was near the co-utilization boundary, with profitability ratios $\gamma_i/\hat{b}_i$ of 1.04824 and 0.91235, respectively (Table~\ref{tbl:profitability}). The LP predicted that glucose would be preferred, but the indifference-curve slope was close to the budget-line slope, indicating that a small change in the constitutive expression rate $\lambda_{\text{fru}}$ could bring the system to degeneracy and allow simultaneous consumption. At $\lambda_{\text{fru}}=0$ the LP returned a corner solution and predicted diauxie.
Raising the constitutive expression rate to the specified-state degeneracy value ($\lambda_{\text{fru}}^{\star}=0.2392$, the value that equalizes the two profitabilities at the initial substrate concentrations) brought the indifference-curve slope into coincidence with the budget-line slope, placing the pair on the degenerate face where co-utilization is one of the optima. At this point the LP still returned corner solutions rather than an interior allocation. Because each profitability $\gamma_i/\hat{b}_i$ depends on its substrate concentration, this equality holds only at the initial state; once consumption drives the concentrations apart the two returns need not remain equal, so the tie places the pair on a state-dependent switching boundary rather than in persistent equality. The simultaneous consumption seen in the trajectory (Fig.~\ref{fig:glucose-fructose}) arose because the corner alternated rapidly across this boundary, with substrate feedback and the solver's basic-solution selection setting the exact switching sequence, and because the constitutive expression $\lambda_{\text{fru}}$ kept the fructose enzyme present, not because the LP selected an interior point on the degenerate face. Because reaching degeneracy required a constitutive rate far above the near-zero value of the regulated enzymes, this demonstrated the mechanism by which co-utilization can arise in the framework rather than a calibrated fit to the glucose--fructose data.

\begin{figure}[h!]
\begin{center}
\includegraphics[width=0.85\textwidth]{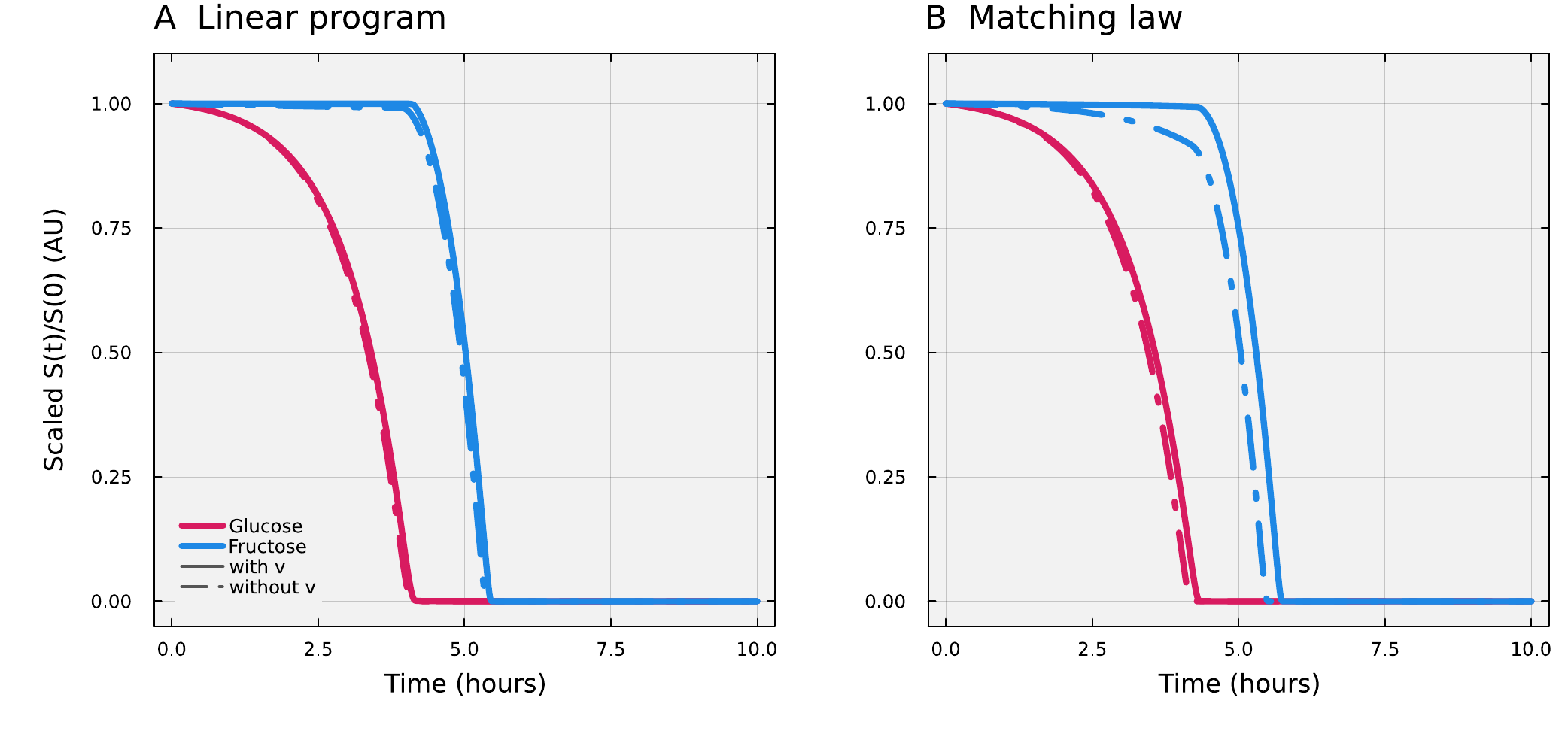}
\end{center}
\caption{Diauxic growth is robust to removing the activity control $v$. Scaled glucose (pink) and fructose (blue) concentrations under the proportional activity law $v_{i}=r_{i}/\max_{k}r_{k}$ (solid) and with it removed, $v_{i}=1$ (dash-dot), overlaid for each synthesis law. \textbf{(A)} Under the linear program the two conditions nearly coincide (co-utilization index $0.005$ with $v$, $0.022$ without). \textbf{(B)} Under the matching law, removing $v$ produces the largest effect, with fructose consumed somewhat earlier (index $0.008$ with $v$, $0.082$ without), though still short of full co-utilization. The co-utilization index is the fraction of fructose consumed before glucose is exhausted.}\label{fig:v-ablation}
\end{figure}

The two cybernetic control variables play distinct roles (Fig.~\ref{fig:v-ablation}). The synthesis control $u$ is the global decision, the corner of the proteome-allocation simplex that commits the entire synthesis budget to a single enzyme and so, through the enzyme it builds, sets which substrate the cell consumes. The activity control $v$ adjusts only the activity of the enzymes already present, a local modulation that leaves the allocation unchanged. To test this, we asked whether diauxie depended on the activity control $v$ or only on the synthesis allocation $u$, using the same near-degenerate glucose--fructose pair. We quantified the outcome with a co-utilization index, the fraction of the less-profitable substrate consumed before the more-profitable one was exhausted. Repeating the simulations with $v_{i}=1$ for all enzymes left diauxie essentially unchanged for glucose--xylose and the triauxic mixture, under both the linear program and the classical matching law, with the substrates still consumed sequentially in the same order (not shown). The activity control was nearly redundant, though for different reasons under the two laws. Under the linear program the synthesis allocation depends only on the profitabilities $\gamma_{i}/\hat{b}_{i}$, set by the substrate concentrations and the fixed cost coefficients and not by the current enzyme abundances; the winner-take-all corner therefore commits synthesis to the most profitable substrate and starves the competing enzyme regardless of its activity. Under the matching law the allocation $u_{i}\propto r_{i}$ does depend on the current enzyme abundance, so synthesis is autocatalytic: the substrate with the higher return accrues synthesis, its enzyme accumulates, and its consumption rate rises further, leaving the competing enzyme scarce. Removing $v$ had a visible effect only for the near-degenerate glucose--fructose pair, and only under the classical matching law, where the co-utilization index rose from 0.008 to 0.082 but still fell short of full co-utilization (Fig.~\ref{fig:v-ablation}B); under the linear program it changed little, from 0.005 to 0.022 (Fig.~\ref{fig:v-ablation}A). Thus, diauxie was a property of the resource-allocation decision rather than of the activity gating.

%% file: sections/discussion.tex
Ramkrishna asked in 1987 whether microbes are optimal strategists~\cite{Ramkrishna:1987aa}. The cybernetic model answered this question by treating enzyme synthesis as the optimal allocation of a limited resource, with the matching rule as the heuristic control law that set the allocation. Young and Ramkrishna~\cite{Young:2007aa} later derived the matching and proportional laws as exact solutions under specific assumptions about the return-on-investment structure of the control problem, and found that the matching law did not maximize average growth rate but reflected a frugal fitness-to-cost tradeoff. That analysis established \emph{that} the matching rule was optimal, but did not identify \emph{what} the cell maximized, nor give a geometric account of diauxie. Here we supplied that missing decision. Casting enzyme allocation as a consumer choice problem, a linear program that maximized a linear growth utility subject to a proteome budget, we found that a linear utility forced the optimum to a corner: the entire budget went to the single most profitable substrate. Thus, diauxie was a corner solution, and the diauxic lag was the time to re-tool synthesis after the program switched corners.

The LP explained the sequential logic of diauxic growth geometrically, at the level of proteome allocation rather than molecular regulation. The dynamic model still contains coarse molecular ingredients, an activity-control law, a substrate-depletion cutoff, enzyme degradation, and prescribed initial enzyme levels, but the decision of which substrate to consume follows from the corner geometry rather than from an explicit catabolite-repression rule. A rational agent maximizing a linear utility subject to a linear budget constraint always chose a corner solution, specializing entirely in the good with the highest utility per unit cost. For a cell, the utility was the growth rate and the cost was the proteome fraction devoted to the synthesis of a particular pseudo enzyme. When glucose supported more growth per unit of proteome investment than xylose, the LP allocated all synthesis resources to the glucose enzyme, $u_{\text{glc}} = 1$, $u_{\text{xyl}} = 0$. This corner persisted until glucose was exhausted, at which point the LP switched to the xylose corner: $u_{\text{glc}} = 0$, $u_{\text{xyl}} = 1$. The diauxic lag was the transient during which the xylose enzyme, previously synthesized at low basal levels, accumulated to the level needed to support rapid growth. Thus, sequential substrate consumption was the dynamic consequence of corner-switching in the LP. Further, the triauxic case (glucose--xylose--lactose) followed without modification: three substrates defined a budget simplex in three dimensions with three corner points. The LP visited them in order of decreasing profitability, producing two diauxic lags and three sequential growth phases.
The linear program and the classical matching law can each be written as a limiting form of a constant-elasticity-of-substitution (CES) utility (Appendix~\ref{sec:ces}). The matching law spread synthesis across substrates in proportion to their returns, the interior optimum of a log (Cobb--Douglas) utility and the diversifying limit ($\sigma\to 1$). By contrast, the linear program committed to a single corner, the optimum of a linear perfect-substitutes utility ($\sigma\to\infty$). The elasticity of substitution is therefore the axis along which the two cybernetic laws differ, distinct from the return-on-investment axis along which Young and Ramkrishna generalized them; because the two limits carry different state-dependent weights, this is a shared limiting form rather than a single fixed-weight, one-parameter family. The two limits also optimized different objectives: the linear program maximized the potential growth return $\sum_i \gamma_i \hat{e}_i$, the return the cell would realize at the target abundances, whereas the matching law optimized the frugal Cobb--Douglas return, so optimality for the linear program meant maximizing the potential return per unit budget at each instant, a greedy optimum, not one proven best over the whole growth trajectory. Whether a cell's effective substitutability is fixed or varies with the substrate pair and environment is a natural question this family raises.

The degenerate case, where two substrates yielded equal profitability, addressed a challenge in the original cybernetic study~\cite{Kompala:1986aa}. At the literature parameters, the linear program ranked glucose just above fructose and predicted sequential use. Thus, the glucose--fructose pair sat near, but not on, the degenerate face. Raising the constitutive expression rate of the fructose enzyme lowered its proteome cost and drove the two ratios together. At the degeneracy point the profitabilities became equal at the initial concentrations, so co-utilization was among the optima there; because the profitabilities are concentration-dependent, this equality marks a state-dependent switching boundary rather than a fixed operating point, and the model produced simultaneous consumption realized through the constitutive fructose expression and rapid corner alternation across that boundary rather than through the LP selecting an interior allocation, reproducing without a separate rule the simultaneous use the original model treated with a modified strategy. Because reaching degeneracy required a constitutive rate above the near-zero value estimated for the regulated enzymes, this was a demonstration of the mechanism by which co-utilization could arise in the utility framework, not a fitted account of glucose--fructose growth. That the linear program and the matching law reproduced the cellmass trajectories almost identically was the expected outcome, and it reinforced the central result: diauxie followed from the corner-switching geometry and was largely insensitive to the particular synthesis law. The two laws diverged only in the allocation domain, where the linear program switched through sharp corners while the matching law reallocated gradually. This difference lived in the synthesis-allocation variables $u_{i}$, the proteome fraction committed to each pseudo enzyme, and was nearly invisible in bulk growth data, so distinguishing the two laws would require allocation-level measurement, such as time-resolved proteomics or a synthesis reporter, rather than a growth curve. Only at the degenerate boundary, which constitutive expression could bring a pair to, did both laws predict co-utilization.

The LP formulation of the cybernetic model is related to flux balance analysis (FBA)~\cite{Orth:2010aa}, an important constraint-based modeling framework in systems biology. Both approaches cast cellular decision-making as an LP, but they operate at different levels of biological organization. Flux balance analysis maximizes a cellular objective (typically growth rate) subject to stoichiometric and thermodynamic constraints on reaction fluxes, operating at the level of metabolic fluxes within a single (potentially genome-scale) metabolic network. On the other hand, the cybernetic LP maximizes growth utility subject to a proteome allocation constraint on enzyme synthesis, operating at the level of resource investment across competing metabolic programs. The closest geometric parallel between the two is phenotype phase-plane analysis~\cite{Edwards:2002aa}, which partitions a two-substrate plane into discrete phases and classifies them by ratios of dual shadow prices at the flux level. The consumer-choice geometry developed here is primal rather than dual, drawn over enzyme abundances as indifference curves against a budget line, and it operates at the level of proteome allocation. Dynamic flux balance analysis extends FBA to substrate switching by solving repeated flux-level linear programs~\cite{Mahadevan:2002aa}, one level below the enzyme-synthesis allocation the cybernetic LP performs. The cybernetic control ideas have already been integrated with structured metabolic networks, most directly in hybrid cybernetic models that overlay cybernetic variables on a flux-mode representation of the network~\cite{Young:2008aa,Kim:2008aa,Vilkhovoy:2016aa}, and more broadly in phenotype prediction~\cite{Varner:2000aa} and the metabolic engineering of biosynthetic pathways~\cite{Varner:1999aa}. Extending this integration to a proteome-allocation linear program over genome-scale stoichiometry (FBA) is a natural direction for future work.

Several simplifications in this formulation should be explored in future work. The utility and budget coefficients $\gamma_i$ and $\hat{b}_i$ came from a linearized steady-state analysis of the enzyme balance equations, and a derivation from first principles would put them on firmer ground. The model used a common enzyme time constant $\tau$ estimated from the literature, though substrate-specific induction kinetics could be resolved instead. The proteome budget was a single scalar inequality, whereas the proteome is partitioned among functional sectors (ribosomes, metabolic enzymes, stress-response proteins), and sector-level constraints from recent proteome-allocation models~\cite{Scott:2010aa,Hui:2015aa} could be incorporated. A quantitative theory of the transient enzyme-accumulation kinetics that set the diauxic lag also remains open. In identifying what the cell maximizes, the linear program gives Ramkrishna's optimal strategist a concrete decision: growth-maximizing specialization under a proteome budget, with diauxie as its corner solution.

%% file: sections/backmatter.tex

\paragraph{Conflict of Interest.}
The author declares that the research was conducted without any commercial or financial relationships that could create a conflict of interest.

\paragraph{Author Contributions.}
J.V. directed the study, developed the model and codes, conducted the simulations and analysis,
generated the figures, and wrote the manuscript. The author reviewed and approved the manuscript.

\paragraph{Funding.}
The work described was supported by the Center on the Physics of Cancer Metabolism through Award Number 1U54CA210184-01 from the National Cancer Institute.
The content is solely the author's responsibility and does not necessarily represent the official views of the National Cancer Institute or the National Institutes of Health.

\paragraph{Data Availability.}
Model code is available under an MIT software license from the Varnerlab GitHub repository \url{https://github.com/varnerlab/Kompala-Model-LP-Paper}.
A versioned snapshot of the code and data is archived at Zenodo (DOI to be assigned on acceptance).

%% file: sections/supplement.tex
\setcounter{equation}{0}
\renewcommand{\theequation}{S\arabic{equation}}

\section{The elasticity-of-substitution family}\label{sec:ces}
The linear program of the main text and the classical matching law can each be written as a limiting form of the constant-elasticity-of-substitution (CES) utility, the standard one-parameter family of consumer preferences~\cite{Varian:1992}. The two limits are taken with different state-dependent weights, so they are shared limiting forms of the CES construction rather than two members of one fixed-weight, one-parameter family, as we make explicit below.
Writing the family on the allocation simplex $\left\{u\geq 0,~\sum_{i}u_{i}=1\right\}$, the CES utility over the allocations $u=\left(u_{1},\dotsc,u_{n}\right)$ with positive weights $w_{i}$ and elasticity of substitution $\sigma\in(0,\infty)$ is given by:
\begin{equation}\label{eqn:ces-utility}
U_{\sigma}(u) = \left(\sum_{i\in\mathcal{S}} w_{i}\,u_{i}^{\alpha}\right)^{1/\alpha},\qquad \alpha \equiv \frac{\sigma-1}{\sigma}.
\end{equation}
The single parameter $\sigma$ sets how readily the cell trades one enzyme allocation for another.
A large $\sigma$ makes the substrates near-perfect substitutes and straightens the indifference curves, while $\sigma\rightarrow 1$ makes them diversifying in the Cobb--Douglas sense.
The weights enter the maximizer only up to a positive rescaling, so we normalize them to $\sum_{i}w_{i}=1$ without changing the optimal allocation.
We show that the limit $\sigma\rightarrow\infty$ recovers the linear program and the limit $\sigma\rightarrow 1$ recovers the matching law. Because the two limits are taken with different weights, the elasticity of substitution is the axis along which the laws differ, but they are not two points of a single fixed-weight, one-parameter CES family.

Consider first the perfect-substitutes limit $\sigma\rightarrow\infty$, where the exponent $\alpha=(\sigma-1)/\sigma\rightarrow 1$.
Both the inner and outer exponents of Eqn.~\eqref{eqn:ces-utility} approach unity, so the utility reduces to the linear form given by:
\begin{equation}\label{eqn:ces-linear-limit}
U_{\sigma}(u) = \left(\sum_{i\in\mathcal{S}} w_{i}\,u_{i}^{\alpha}\right)^{1/\alpha}\;\xrightarrow[\alpha\rightarrow 1]{}\;\sum_{i\in\mathcal{S}} w_{i}\,u_{i}.
\end{equation}
Setting the weights to the profitabilities, $w_{i}=\rho_{i}=\gamma_{i}/\hat{b}_{i}$ from Eqn.~\eqref{eqn:profitability-def}, gives the objective $\sum_{i}\rho_{i}u_{i}$ of the linear program.
The maximum of this linear utility over the simplex is attained at the corner $i^{\star}=\arg\max_{i}\rho_{i}$, the perfect-substitutes optimum that produces diauxic specialization.

Consider next the diversifying limit $\sigma\rightarrow 1$, where $\alpha\rightarrow 0$.
Because a monotone transformation leaves the maximizer unchanged, we take the logarithm and expand $u_{i}^{\alpha}=\exp(\alpha\ln u_{i})=1+\alpha\ln u_{i}+O(\alpha^{2})$.
With the weights normalized to $\sum_{i}w_{i}=1$, the sum inside the logarithm is $1+\alpha\sum_{i}w_{i}\ln u_{i}+O(\alpha^{2})$, which gives:
\begin{equation}\label{eqn:ces-cobb-limit}
\ln U_{\sigma}(u) = \frac{1}{\alpha}\ln\!\left(\sum_{i\in\mathcal{S}} w_{i}\,u_{i}^{\alpha}\right) = \frac{1}{\alpha}\ln\!\left(1+\alpha\sum_{i\in\mathcal{S}} w_{i}\ln u_{i}+O(\alpha^{2})\right)\;\xrightarrow[\alpha\rightarrow 0]{}\;\sum_{i\in\mathcal{S}} w_{i}\ln u_{i},
\end{equation}
where the last step uses $\ln(1+x)=x+O(x^{2})$.
Exponentiating recovers the Cobb--Douglas utility $\prod_{i}u_{i}^{w_{i}}$ of Eqn.~\eqref{eqn:cobb-douglas-utility}.
Setting the weights to the substrate returns, $w_{i}=r_{i}$, gives the objective $\sum_{i}r_{i}\ln u_{i}$ of the matching law, whose maximizer is the interior allocation $u_{i}=r_{i}/\sum_{k}r_{k}$ of Eqn.~\eqref{eqn:matching-law}.

The family therefore places the two cybernetic laws at opposite ends of a single axis, the elasticity of substitution $\sigma$: the linear program is the perfect-substitutes end ($\sigma\rightarrow\infty$) and the matching law is the diversifying end ($\sigma\rightarrow 1$).
The two limits carry different state-dependent weights, the profitabilities $\rho_{i}$ for the linear program and the returns $r_{i}$ for the matching law, so the family relates the laws along substitutability rather than as a single fixed-weight interpolation.
This axis is distinct from the generalization of Young and Ramkrishna, who varied the return-on-investment structure within the interior allocation while leaving the diversifying character of the matching law intact~\cite{Young:2007aa}.

%% file: References_v1.bib
@article{Kompala:1986aa,
	author = {Kompala, D S and Ramkrishna, D and Jansen, N B and Tsao, G T},
	doi = {10.1002/bit.260280715},
	journal = {Biotechnol Bioeng},
	journal-full = {Biotechnology and bioengineering},
	month = {Jul},
	number = {7},
	pages = {1044-55},
	pmid = {18555426},
	pst = {ppublish},
	title = {Investigation of bacterial growth on mixed substrates: experimental evaluation of cybernetic models},
	volume = {28},
	year = {1986}}

@article{Ramkrishna:1987aa,
	author = {Ramkrishna, D and Kompala, D S and Tsao, G T},
	title = {Are Microbes Optimal Strategists?},
	journal = {Biotechnol Prog},
	volume = {3},
	number = {3},
	pages = {121--126},
	year = {1987},
	doi = {10.1002/btpr.5420030302}}

@article{Dhurjati:1985aa,
	author = {Dhurjati, P and Ramkrishna, D and Flickinger, M C and Tsao, G T},
	title = {A cybernetic view of microbial growth: modeling of cells as optimal strategists},
	journal = {Biotechnol Bioeng},
	volume = {27},
	pages = {1--9},
	year = {1985},
	doi = {10.1002/bit.260270102}}

@article{Kompala:1984aa,
	author = {Kompala, D S and Ramkrishna, D and Tsao, G T},
	title = {Cybernetic modeling of microbial growth on multiple substrates},
	journal = {Biotechnol Bioeng},
	volume = {26},
	number = {11},
	pages = {1272--1281},
	year = {1984},
	doi = {10.1002/bit.260261103}}

@incollection{Ramkrishna:1982,
	author = {Ramkrishna, D},
	title = {A Cybernetic Perspective of Microbial Growth},
	booktitle = {Foundations of Biochemical Engineering: Kinetics and Thermodynamics in Biological Systems},
	publisher = {American Chemical Society},
	address = {Washington, DC},
	series = {ACS Symposium Series},
	volume = {207},
	pages = {161--178},
	year = {1983},
	doi = {10.1021/bk-1983-0207.ch007}}

@book{Monod:1942,
	author = {Monod, Jacques},
	title = {Recherches sur la Croissance des Cultures Bact\'{e}riennes},
	publisher = {Hermann et Cie},
	address = {Paris},
	year = {1942}}

@article{Deutscher:2008,
	author = {Deutscher, Josef},
	title = {The mechanisms of carbon catabolite repression in bacteria},
	journal = {Curr Opin Microbiol},
	volume = {11},
	number = {2},
	pages = {87--93},
	year = {2008},
	doi = {10.1016/j.mib.2008.02.007}}

@book{Varian:1992,
	author = {Varian, Hal R},
	title = {Microeconomic Analysis},
	publisher = {W. W. Norton \& Company},
	address = {New York},
	edition = {3rd},
	year = {1992}}

@article{Orth:2010aa,
	author = {Orth, Jeffrey D and Thiele, Ines and Palsson, Bernhard {\O}},
	title = {What is flux balance analysis?},
	journal = {Nat Biotechnol},
	volume = {28},
	number = {3},
	pages = {245--248},
	year = {2010},
	doi = {10.1038/nbt.1614}}

@article{Mahadevan:2002aa,
	author = {Mahadevan, Radhakrishnan and Edwards, Jeremy S and Doyle, III, Francis J},
	title = {Dynamic flux balance analysis of diauxic growth in {Escherichia coli}},
	journal = {Biophys J},
	volume = {83},
	number = {3},
	pages = {1331--1340},
	year = {2002},
	pmid = {12202358},
	doi = {10.1016/S0006-3495(02)73903-9}}

@article{Edwards:2002aa,
	author = {Edwards, Jeremy S and Ramakrishna, Ramprasad and Palsson, Bernhard O},
	title = {Characterizing the metabolic phenotype: a phenotype phase plane analysis},
	journal = {Biotechnol Bioeng},
	volume = {77},
	number = {1},
	pages = {27--36},
	year = {2002},
	pmid = {11745171},
	doi = {10.1002/bit.10047}}

@article{Scott:2010aa,
	author = {Scott, Matthew and Gunderson, Carl W and Mateescu, Eduard M and Zhang, Zhongge and Hwa, Terence},
	title = {Interdependence of cell growth and gene expression: origins and consequences},
	journal = {Science},
	volume = {330},
	number = {6007},
	pages = {1099--1102},
	year = {2010},
	doi = {10.1126/science.1192588}}

@article{Hui:2015aa,
	author = {Hui, Sheng and Silverman, Josh M and Chen, Stephen S and Erickson, David W and Basan, Markus and Wang, Jilong and Hwa, Terence and Williamson, James R},
	title = {Quantitative proteomic analysis reveals a simple strategy of global resource allocation in bacteria},
	journal = {Mol Syst Biol},
	volume = {11},
	number = {784},
	year = {2015},
	doi = {10.15252/msb.20145697}}

@article{Young:2015aa,
	author = {Young, Jamey D},
	title = {Learning from the Steersman: A Natural History of Cybernetic Models},
	journal = {Ind Eng Chem Res},
	volume = {54},
	number = {42},
	pages = {10162--10169},
	year = {2015},
	doi = {10.1021/acs.iecr.5b01315}}

@article{Straight:1994aa,
	author = {Straight, Jeffrey V and Ramkrishna, Doraiswami},
	title = {Cybernetic modeling and regulation of metabolic pathways. {G}rowth on complementary nutrients},
	journal = {Biotechnol Prog},
	volume = {10},
	number = {6},
	pages = {574--587},
	year = {1994},
	doi = {10.1021/bp00030a002}}

@article{Young:2008aa,
	author = {Young, Jamey D and Henne, Kristene L and Morgan, John A and Konopka, Allan E and Ramkrishna, Doraiswami},
	title = {Integrating cybernetic modeling with pathway analysis provides a dynamic, systems-level description of metabolic control},
	journal = {Biotechnol Bioeng},
	volume = {100},
	number = {3},
	pages = {542--559},
	year = {2008},
	doi = {10.1002/bit.21780}}

@article{Varner:1999aa,
	author = {Varner, Jeffrey D and Ramkrishna, Doraiswami},
	title = {Metabolic engineering from a cybernetic perspective: aspartate family of amino acids},
	journal = {Metab Eng},
	volume = {1},
	number = {1},
	pages = {88--116},
	year = {1999},
	pmid = {10935757},
	doi = {10.1006/mben.1998.0104}}

@article{Varner:2000aa,
	author = {Varner, Jeffrey D},
	title = {Large-scale prediction of phenotype: concept},
	journal = {Biotechnol Bioeng},
	volume = {69},
	number = {6},
	pages = {664--678},
	year = {2000},
	pmid = {10918142},
	doi = {10.1002/1097-0290(20000920)69:6<664::aid-bit11>3.0.co;2-h}}

@article{Kim:2008aa,
	author = {Kim, Jin Il and Varner, Jeffrey D and Ramkrishna, Doraiswami},
	title = {A hybrid model of anaerobic {E. coli} {GJT001}: combination of elementary flux modes and cybernetic variables},
	journal = {Biotechnol Prog},
	volume = {24},
	number = {5},
	pages = {993--1006},
	year = {2008},
	pmid = {19194908},
	doi = {10.1002/btpr.73}}

@article{Vilkhovoy:2016aa,
	author = {Vilkhovoy, Michael and Minot, Mason and Varner, Jeffrey D},
	title = {Effective dynamic models of metabolic networks},
	journal = {IEEE Life Sci Lett},
	volume = {2},
	number = {4},
	pages = {51--54},
	year = {2016},
	doi = {10.1109/LLS.2016.2644649}}

@book{RamkrishnaSong:2018aa,
	author = {Ramkrishna, Doraiswami and Song, Hyun-Seob},
	title = {Cybernetic Modeling for Bioreaction Engineering},
	publisher = {Cambridge University Press},
	address = {Cambridge, UK},
	series = {Cambridge Series in Chemical Engineering},
	year = {2018},
	doi = {10.1017/9780511731969}}

@article{Young:2007aa,
	author = {Young, Jamey D and Ramkrishna, Doraiswami},
	doi = {10.1021/bp060176q},
	journal = {Biotechnol Prog},
	journal-full = {Biotechnology progress},
	mesh = {Algorithms; Bacterial Physiological Phenomena; Cell Proliferation; Cybernetics; Models, Biological; Monte Carlo Method},
	number = {1},
	pages = {83-99},
	pmid = {17269675},
	pst = {ppublish},
	title = {On the matching and proportional laws of cybernetic models},
	volume = {23},
	year = {2007}}

@article{Bezanson:2017aa,
	author = {Bezanson, Jeff and Edelman, Alan and Karpinski, Stefan and Shah, Viral B},
	title = {Julia: A Fresh Approach to Numerical Computing},
	journal = {SIAM Rev},
	volume = {59},
	number = {1},
	pages = {65--98},
	year = {2017},
	doi = {10.1137/141000671}}

@article{Lubin:2023aa,
	author = {Lubin, Miles and Dowson, Oscar and Dias Garcia, Joaquim and Huchette, Joey and Legat, Beno{\^i}t and Vielma, Juan Pablo},
	title = {{JuMP} 1.0: Recent improvements to a modeling language for mathematical optimization},
	journal = {Math Program Comput},
	volume = {15},
	pages = {581--589},
	year = {2023},
	doi = {10.1007/s12532-023-00239-3}}

@misc{GLPK,
	author = {Makhorin, Andrew},
	title = {{GLPK}: {GNU} Linear Programming Kit},
	howpublished = {\url{https://www.gnu.org/software/glpk/}}}
